\documentclass[acp,nonumber]{copernicus}

\usepackage{amssymb}
\usepackage{amsmath}
\usepackage{bm}
\usepackage{graphicx}					% Include figure files
\usepackage{version}				
\usepackage{cases}						%Pour les systmes d'quations

\usepackage{mathrsfs}

\usepackage{graphicx}
\usepackage{color}
\usepackage{psfrag}
\usepackage{enumerate}
\usepackage{amssymb}
\usepackage{amsmath}
\usepackage{wasysym}
\usepackage{epstopdf}	
\usepackage{lipsum, babel}

\usepackage[dvipsnames]{xcolor}

\newcommand{\ew}[1]{\textcolor{black}{#1}}
\newcommand{\ews}[1]{\textcolor{black}{#1}}

\begin{document}
\nolinenumbers

\title{Intricate Relations Among Particle Collision, Relative Motion and Clustering in Turbulent Clouds: Computational Observation and Theory.
%Nontrivial Causality and Theory of Particle Collision, Relative Motion and Clustering in Turbulent Cloud.
}

% \Author[affil]{given_name}{surname}

\Author[1,2]{Ewe-Wei Saw}{}
\Author[1]{Xiaohui Meng}{}
%\Author[]{}{}

\affil[1]{School of Atmospheric Sciences and Guangdong Province Key Laboratory for Climate Change and Natural Disaster Studies, Sun Yat-Sen University, Zhuhai, China}
\affil[2]{Ministry of Education Key Laboratory of Tropical Atmosphere-Ocean System, Zhuhai, China }

%% The [] brackets identify the author with the corresponding affiliation. 1, 2, 3, etc. should be inserted.

%% If an author is deceased, please mark the respective author name(s) with a dagger, e.g. "\Author[2,$\dag$]{Anton}{Smith}", and add a further "\affil[$\dag$]{deceased, 1 July 2019}".

%% If authors contributed equally, please mark the respective author names with an asterisk, e.g. "\Author[2,*]{Anton}{Smith}" and "\Author[3,*]{Bradley}{Miller}" and add a further affiliation: "\affil[*]{These authors contributed equally to this work.}".

\correspondence{E.-W. Saw (ewsaw3@gmail.com), X. Meng (mengxh7@mail2.sysu.edu.cn) }

\runningtitle{ }

\runningauthor{E.-W. Saw and X. Meng}

\received{}
\pubdiscuss{} %% only important for two-stage journals
\revised{}
\accepted{}
\published{}

%% These dates will be inserted by Copernicus Publications during the typesetting process.

\firstpage{1}

\maketitle

\begin{abstract}

\ew{Considering turbulent clouds containing small inertial particles, we investigate the effect of particle collision, in particular collision-coagulation, on particle clustering and particle relative motion. We perform direct numerical simulation (DNS) of coagulating particles in isotropic turbulent flow in the regime of small Stokes number ($St=0.001-0.54$) and find that, due to collision-coagulation, the radial distribution functions (RDFs) fall-off dramatically at scales $r \sim d\,\,$ (where $d$ is the particle diameter) to small but finite values, while the mean radial-component of particle relative velocities (MRV) increase sharply in magnitudes. Based on a previously proposed Fokker-Planck (drift-diffusion) framework, we derive a theoretical account of the relationship among particle collision-coagulation rate, RDF and MRV. The theory includes contributions from turbulent-fluctuations absent in earlier mean-field theories. We show numerically that the theory accurately accounts for the DNS results (i.e., given an accurate RDF, the theory could produce an accurate MRV). Separately, we also propose a phenomenological model that could directly predict MRV and find that it is accurate when calibrated using fourth moments of the fluid velocities. We use the model to derive a general solution of RDF. We uncover a paradox: the past empirical success of the differential version of the theory is theoretically unjustified. We see a further shape-preserving reduction of the RDF (and MRV) when the gravitational settling parameter ($S_g$) is of order $O(1)$. Our results demonstrate strong coupling between RDF and MRV and imply that earlier isolated studies on either RDF or MRV have limited relevance for predicting particle collision rate.}

\end{abstract}

%%\copyrightstatement{TEXT} %% This section is optional and can be used for copyright transfers.

\introduction  %% \introduction[modified heading if necessary]
The motion and interactions of small particles in turbulence have fundamental implications for atmospheric clouds; specifically, they are relevant to the time-scale of rain formation particularly in warm-clouds  \citep{Falkovich02, Wilkinson06, Grabowski13} [a similar problem also applies to planet formation in astrophysics  \citep{Johansen07}]. They are also important for engineers who are designing future, greener, combustion engines, as this is a scenario they wish to understand and control in order to increase fuel-efficiency \citep{Karnik12}. Cloud particles or droplets, due to their inertia, are known to be ejected from turbulent vortices and thus form clusters -- regions of enhanced particle-density \citep{wood05, Bec07, Saw08, Karpinska19}; this together with droplet collision is of direct relevance for 
the mentioned applications. Due to the technical difficulty of obtaining extensive and systematic experimental or field data on particle/droplet collision in turbulent cloud, many of the recent studies rely on direct numerical simulation (DNS), examples of which could be found in, e.g., \citep{Onishi16, Wang08} and reference therein.
%burning efficiency of combustion engines\cite{Karnik12}. 
%Beside these, particle-turbulence interaction also plays a significant role in industrial reactors and nanoparticle generation\cite{Pratsinis1996}. 
Up until now, we do not have definitive answers to basic questions such as how to calculate particle collision rate from basic turbulence-particle parameters and what is the exact relation between collision and particle clustering and/or motions, for, as we shall see, our work reveals that collision-coagulation causes profound changes in \ew{particle relative velocity statistics and particle clustering, questioning earlier understanding of the problem.} The difficulty of this problem is in part related to the fact that turbulence is, even by itself, virtually intractable theoretically due to its nonlinear and complex nature. 

%However, as we shall see, accurate reduced models could sometimes be afforded with the help of methods like the Fokker-Planck equation \cite{Risken1996}.

The quest for a theory of particle collision in turbulence started in 1956 when %Saffman and Turner 
\cite{Saffman1956} derived a mean-field formula
%in the seminar work of Saffman \& Turner \cite{Saffman1956} derived a formula 
for collision rate of finite size, inertialess, particles. In another landmark work \citep{Sundaram1997}, a general relation among collision-rate ($R_c$), particle clustering and mean particle relative radial velocity was presented: \ews{$R_c / (n_1 n_2 V) \!=\!  4\pi d^2  g(d) \left< w_r(d) \right>_{\!*}\,$, } %(hereafter: S97) 
%$R_c / (n_1 n_2 V) \!=\!  4\pi d^2  g(d) \left< w_r(d) \ | w_r \le 0 \right>_{*}\,$,
where $g(r)$ is the particle radial distribution function (RDF), %a metric of the degree of particle clustering, 
$w_r$ is the radial component of relative velocity between two particles, $\left< \cdot \right>_{\!*}$ denotes averaging over particle-pairs, $\left< w_r(d) \right>_{\!*}$ is the mean radial-component of relative particle velocity (MRV)%\footnote{Note: the condition $w_r \le 0$ is needed in the calculation of MRV in \citep{Sundaram1997} because they were considering "ghost-particles" that are non-colliding, since without that condition, MRV is always zero in turbulence. In our work, such conditioning is both unnecessary and incorrect.}, } 
, $n_i$'s are \ew{global averages of} particle number density, $V$ is the spatial volume of the domain, \ew{$d$ the particle diameter}. The remarkable simplicity of this finding inspired a "separation paradigm", %whereby many subsequence studies focused either solely on the RDF %(e.g. studies involving collision-less `ghost' particles)  
%or  solely on MRV, often with anticipation that their findings, besides being of scientific significance by its own right, would eventually contribute to the prediction of $R_c$. 
which \ew{is the idea} that one could study the RDF or MRV separately (which \ew{are} technically easier), the independent results from the \ews{two} may be combined to accurately predicts $R_c$ (an idea that we subsequently challenge).
Another work of special interest here is the drift-diffusion model by \cite{Chun05} (hereafter: CK theory) 
 (note: there are other \ew{similar} theories \citep{Balkovsky01, Zaichik03}). The CK theory, derived for non-colliding particles in the limit of vanishing particle Stokes number $St$ (a quantity that reflects the importance of the particle's inertia in dictating its motion in turbulence), correctly predicted the power-law form of the RDF \citep{Reade00,Saw08} and have seen remarkable successes over the years including the accurate account of the modified RDF of particles interacting electrically \citep{Lu10} and hydrodynamically \citep{Yavuz18}.

Here, we first present results on RDF and MRV for particles undergoing collision-coagulation\footnote{Coagulation is, in a sense, the simplest outcome of collision. In the sequel we shall argue that the major qualitative conclusions of our work also apply to cases with other collisional outcomes.}. 
The data is obtained via direct numerical simulation (DNS), which is the gold-standard computational method in terms of accuracy and completeness for solving the most challenging fluid dynamics problem, i.e., turbulent flows. %DNS solves the fundamental equation of fluid dynamics, the Navier-Stokes Equation, with full resolution and without turbulence modeling. %, a feat that analogs the ab-initio methods in quantum-physics/chemistry with respect to the Schr{\"o}dinger equation. 
%The accuracy of DNS for various turbulent-flows have been experimentally validated for decades (see e.g. the compilation of results in \cite{Pope00}); while for simulating dynamics of small inertial particles, experimental validation of its accuracy could be found in \cite{Salazar08, Saw12b, Saw14, Dou18}. 
\ew{It is worth noting that the focus of our work is on the fundamental relationship between collision, RDF and MRV, and to highlight differences from the case with non-colliding particles \citep{Chun05}. To that end, we have designed the DNS to have an idealized setup similar to what was done in \citep{Chun05}, which would allow us to identify without doubt the effects of particle collision-coagulation. As a result, this limits the direct applicability to real systems (these limitations are detailed in Sec.~\ref{Limitations}).} 

Analysis of the DNS results is followed by a theoretical account of the relations between collision-rate, RDF and MRV which includes mean-field contributions \citep{Saffman1956, Sundaram1997} and contributions from turbulent fluctuations (absent from earlier theories \citep{Saffman1956, Sundaram1997}). 
%We show that such a such complete treatment of the collision problem can be 
The theory is derived from the Fokker-Planck (drift-diffusion) framework first introduced in the CK theory \citep{Chun05}.
We shall see that the main effect of collision-coagulation is the enhanced asymmetry in the particle relative velocity distribution\footnote{In the collision less case, the asymmetry is much weaker and is related to viscous dissipation of energy in turbulence \citep{Pope00}.} and that this leads to nontrivial outcomes. %We challenge the "separation paradigm" by showing that collision-with-coagulation leads to profoundly different RDF and MRV; and to strong coupling between RDF and MRV; such that results from any studies that preclude particle collision has limited relevance for predicting collision statistics\footnote{The later part of the the current statement still holds with other types of collisional outcomes, but the results should be qualitatively different from the case of collision-coagulation.}. %It is also evident that there is a strong coupling between RDF and MRV when collisions occurs. %(if not in general). 

\section{Direct Numerical Simulation (DNS)}
%\label{methods}

%\subsection{Brief overview of the direct numerical simulation}
%We perform DNS to observe how collision-coagulation affects RDF and MRV. 
%We now briefly describe the DNS. It is a standard pseudo-spectral algorithm \citep{Rogallo1981, Pope00, Mortensen16} that solves the Navier-Stokes Equations with brute force giving fully resolved isotropic turbulent flow field (more in "Methods"). The flow has Taylor-scaled Reynolds number $R_{\lambda}=133$. The particles are moved via Stokes drag: $\dot{\bm{v}} = (\bm{u}-\bm{v})/\tau_p$, where $\bm{u}, \bm{v}$ is the local fluid and particle velocity, $\tau_p$ is the particle inertia respond time. Exponential integrator method \citep{Ireland13} is employed to give accurate particle trajectories even at very small $St$, (note: $St=\frac{1}{18}(\rho_p/\rho)(d/\eta)^2$, where $\rho_p/\rho$ is the particle-to-fluid mass-density ratio, $d$ is the particle diameter, $\eta$ the Kolmogorov length-scale). The particles collide when their (spherical) volume overlap and a new particle is formed conserving volume and momenta. We continuously, randomly, inject new particles into the flow so that the system is in a steady-state after some time. Statistical analyses are done at steady-state on monodisperse particles (involving particles with the same $St$).

%\subsection{Details of the Direct Numerical Simulation.}
To observe how particle collision-coagulation affects RDF and MRV, we performed direct numerical simulation (DNS) of steady-state isotropic turbulence embedded with particles of finite but sub-Kolmogorov size. We solve the incompressible Navier-Stokes Equations (Eq.~(\ref{NS_eqn})) using the standard pseudo-spectral method \citep{Rogallo1981, Pope00, Mortensen16} inside a triply periodic cubic-box.% of size $2\pi \times\!\! 10^{-\!1}m$. 
\begin{multline} \label{NS_eqn}
\ \ \ \ \ \ \ \ \ \   \frac{\partial \mathbf{u} }{\partial t} + \mathbf{u} \cdot \nabla \mathbf{u} = -  \frac{1}{\rho} \nabla p + \nu\nabla^2 \mathbf{u} + \mathbf{f}(\mathbf{x},t)  \,, \\
\nabla \cdot \mathbf{u} = 0\,, \ \ \ \ \ \ \ \ \ \ \ \ \  \ \ \ \ \ \ \ \ \ \ \ \ \ \ \ \ \ \ \ \ \ \ \ \ \ \ \ \ \ \ \ \
\end{multline}
where $\rho,\, p, \,\nu,\, \mathbf{f}$ are the fluid mass-density, pressure, kinematic viscosity, imposed forcing respectively. The velocity field is discretized on a $256^3$ grid. 
 %Efficient calculation is afforded by computing the linear (nonlinear) terms of the equation in Fourier space (real space). 
 Aliasing resulting from Fourier transform of truncated series is removed via a $2/3$-dealiasing rule \citep{Rogallo1981}. 
 %so that the maximum wave-number in our simulations is $k_{max} = N/3$. 
 A statistically stationary and isotropic turbulent flow is achieved by continuously applying random forcing to the lowest wave-numbers until the flow's energy spectrum is in steady-state \citep{Eswaran1988}. %yielding a Taylor-scaled Reynolds number $R_{\lambda}=133$. 
 The 2nd-order Runge-Kutta time stepping was employed. Further details of such a standard turbulence simulator can be found in, e.g.,~\cite{Pope00}, \cite{Rogallo1981} and \cite{Mortensen16}. 
The accuracy of DNS for turbulent flows have been experimentally validated for decades (see, e.g., the compilation of results in \citep{Pope00}). %while for simulating dynamics of small inertial particles, experimental validation of its accuracy could be found in \cite{Salazar08, Saw12b, Saw14, Dou18}. 
 
Particles in the simulations are advected via a viscous Stokes drag force \ew{\citep{Maxey1983}}:
\[
d\mathbf{v}/dt = (\mathbf{u}-\mathbf{v})/\tau_p\,, 
\]
where $\mathbf{u}, \mathbf{v}$ \ew{are} the local fluid and particle velocity respectively, $\tau_p$ is the particle inertia response time, \ew{defined as $\tau_p= \frac{1}{18}(\rho_p/\rho -1)(d^2/\nu)$}, where $\rho_p$ is the particle mass-density and $d$ is the particle diameter. 
%The focus of our study is on the effect of particle collision-coagulation in the simplest fluid dynamic setting so that its implication and fundamental interaction with turbulence can be fully understood before moving on to more complex settings in future works. 
\ew{As mentioned, this work focuses on the fundamental relationship between collision-coagulation, RDF and MRV, as well as on addressing the validity of the theory (to be described). It is thus, beneficial to keep the DNS setting idealized (and in the regime relevant for the theory) for the sake of clarity when interpreting results. To that end, the DNS does not include inter-particle hydrodynamic interactions (HDI) and gravitational settling, nor does it consider the effects of temperature-, humidity-variation and phase transitions. Such practice is not uncommon in studies designed to isolate and address fundamental issues related to particles dynamics in turbulence, examples that are closely related to the current setup and/or problem include \citep{Sundaram1997, Chun05, Bec07, Salazar08, Wang08, Woittiez09, Vosskuhle13}. However, such an approach certainly limits the direct applicability of our results to some realistic problems in  the atmosphere, these limitations will be detailed in Sec.~\ref{Limitations}, where a discussion of the effects of gravity and HDI is also given.} 
%(this implies that if practical applicability is of concern, the current results only apply to cloud particles with gravitational terminal velocities that are small compared to the velocity scale of the smallest turbulent eddies, e.g. particle of size $\lesssim 50 \mu m$ in atmospheric clouds). %(note also that the only viable theoretical treatment of this problem that we shall introduce in sequel assumes the asymptotic of very small particle which in turn implies limited importance of graviational setlling). 

In this context, the particle Stokes number, defined as $\tau_p/\tau_{\eta}$ where $\tau_{\eta}$ is the Kolmogorov time-scale, could be expressed as $St=\frac{1}{18}(\rho_p/\rho -1)(d/\eta)^2$, where $\eta$ is the Kolmogorov length-scale. Time-stepping of the particle motion is done using a 2nd-order modified Runge-Kutta method with "exponential integrator" that is accurate even for $\tau_p$ much smaller than the fluid's time-step \citep{Ireland13}. \ew{The particles introduced into the simulation are spherical and are of the same size, the initial number of particles is $10^7$ and they are randomly distributed in space. Particles} collide when their volumes overlap and a new particle is formed conserving volume and momentum \citep{Bec16}. %\footnote{they are essentially tiny droplets with infinite surface tension.}. 
We continuously, randomly, inject new particles so that the system is in a steady-state after some time. Statistical analysis is done at steady-state on monodisperse particles (i.e., particles with the same $St$). \ew{Experimental validation of the  accuracy of such particle simulating scheme in DNS could be found in \cite{Salazar08, Saw12b, Saw14, Dou18}.}

Values of key parameters of the DNS are given in Table~\ref{table1}. \ew{Values of other parameters and further details could be found in \citep{Supp}.}

% Key parameters of the DNS: $R_{\lambda}=133$; kinetic energy dissipation rate, $\varepsilon=0.117$; fluid kinematic viscosity, $\nu=0.001$; Kolmogorov lengh-scale, $\eta=0.00962$; Kolmogorov time-scale, $\tau_{\eta}=0.0925$. Initial diameter of the particles is $d_*\!=\!9.49\!\times\! 10^{-4}$ unless otherwise specified (in one case, particles are initialized at $2d_*$ in order to observe the effect of size variation). \ew{ As discussed above, particle Stokes number is defined as: $St \equiv \tau_p/\tau_{\eta} \equiv \frac{1}{18}(\rho_p/\rho)(d/\eta)^2$.} %where $\rho_p/\rho$ is the particle to fluid mass-density ratio, $d$ is the particle diameter.

\begin{center} 
\begin{table*}[!] 
\begin{tabular}{c c c c c c c c c} 
\hline
$Re_{\lambda}$ & $\nu \ [ \text{dm}^{2}\text{/s}]$ & $u_{rms} \,[\text{dm/s}]$  & $\epsilon \ [\text{dm}^{2}\text{/s}^{3}]$ & $\eta \ [\text{dm}]$ & $\tau_{\eta} \ [s]$ & $L_c \ [\text{dm}]$ & $d \ [\text{dm}]$ \\
%\hline
133 & 0.001 & 0.613 & 0.117 & 0.00962 & 0.0925 & $2\pi$ & $d_* \text{ or } 2d_*$ \\
\hline
\vspace{0pt}
\end{tabular}
\caption{\ew{Values of the parameters in the DNS. (Note: dm~$=$ decimeter). From the left,} we have the Taylor-scale Reynolds number, kinematic viscosity of the fluid, root-mean-square of fluid velocity, kinetic energy dissipation rate, Kolmogorov length- and time-scale, length of the simulation cube and particle diameters considered. We have introduced $d_*$ to represent the specific value: \ew{$\!9.49\!\times\! 10^{-4} \text{dm}$ (more details in the text)}. We choose the units of the length (time) scale in the DNS to be in decimeter (second), such that $\nu$ is nearly its typical value in the atmosphere.} \label{table1}
\end{table*}
\end{center}

  \section{Elements of the Drift-Diffusion Theory} %Origin of the Master Equation. in the Manuscript} 
 \label{derive_masterEqn}
 
 \ew{As described in \citep{Chun05}, in the limit of} $St \ll 1$, particle motions are closely tied to the fluid \ew{velocity} and, to leading order, completely specified by the particle position and fluid velocity gradients. We consider the Fokker-Planck equation which is closed and deterministic (see, e.g., Appendix J in \citep{Pope00}):
\begin{equation} \label{}
\frac{\partial  P}{\partial t} + \frac{\partial (W_i P)}{\partial r_i} =0 \,,
\end{equation}
where $P \equiv P(r_i, \, t \,\, |\, \Gamma_{ij}(t))$ is the (per volume) probability density (PDF) for a secondary particle %(which could have any history) 
to be at vector position $r_i$ relative to a primary particle at time $t$, conditioned on a fixed and known history of the velocity gradient tensor along \ew{the primary particle's} trajectory $\Gamma_{ij}(t)$, $W_i$ is the mean velocity of secondary particles relative to the primary, under the same \ew{condition. Note: $W_i$ is a conditional-average, while $w_i$ denotes a realization of relative velocity between two particle}. 

From this, one could derive an equation for \ew{ $\left< P \right>\!\,$}:
\begin{equation} \label{}
\frac{\partial \left< P \right>}{\partial t} + \frac{\partial}{\partial r_i} \!\left(\, \left< W_i \right> \! \left< P \right> + \left< W_i P' \right> \,\right) =0 \,,
\end{equation}
\ew{where $\left< . \right>$ implies ensemble averaging over primary particle histories (note: $\left< W_r \right> \equiv$ unconditional mean of $w_r$, averaged over all particle pairs, \ews{i.e., the MRV}). This equation, however, is not closed} due to the correlation between the fluctuating terms $W_i$ and $P' \equiv P -\left< P \right>$. %In our case $\left< W_i \right> \! \left< P \right>$ is not zero.
\ew{The correlation} $\left< W_i P' \right>$ can be written in terms of \ew{a drift flux and diffusive flux (detailed derivation is well described in \citep{Chun05}), such that we have}:
 \begin{equation} \label{master1}
\frac{\partial \left< P \right>}{\partial t} + \frac{\partial}{\partial r_i} \left( q_i^d +q_i^D \right) +\frac{\partial (\left< W_i \right>\left< P \right>)}{\partial r_i} =0 \,,
\end{equation}
where the drift flux is: 
\begin{equation} \label{def_q_d}
 q_i^d \!=\! -\int_{-\infty}^{t}\left< W_i(\mathbf{r},t) \frac{\partial W_l}{\partial r'_l}(\mathbf{r}',t') \right>  \left< P \right> \!(\mathbf{r}',t')\,dt' ,
\end{equation}  
and the diffusive flux is:  
\begin{equation} \label{def_q_D}
	q_i^D \!=\! -\int_{-\infty}^{t}\left< W_i(\mathbf{r},t)  W_j(\mathbf{r}',t') \right> \frac{ \partial \left<P\right>}{\partial r'_j} \!(\mathbf{r}',t')\,dt' ,
\end{equation} 
where $\mathbf{r}'$ satisfies a characteristic equation: 
$ \frac{\partial r'_i}{\partial t'}=W_i(\mathbf{r}',t') \,, $
with boundary condition $r'_i=r_i\,$ at $t'=t$.

\ews{Finally we note that, since particles are allowed to collide-coagulate in our theory, we use the conventional definition of MRV: $\left< W_r \right> \equiv \left< w_r \right>_{\!*}$. In some works that consider non-colliding (ghost) particles, the conditional mean $\left< w_r \ | \ w_r \le 0\right>_{\!*}$ must be used for the purpose of calculating mean collision rate, since there, $\left< w_r \right>_{\!*} = 0$ due to local isotropy of turbulence \citep{Chun05}.} 

\section{DNS Results, Theory and Discussion}  \label{results}
\begin{figure}
  \begin{center}
     {\includegraphics[width=.49 \textwidth]{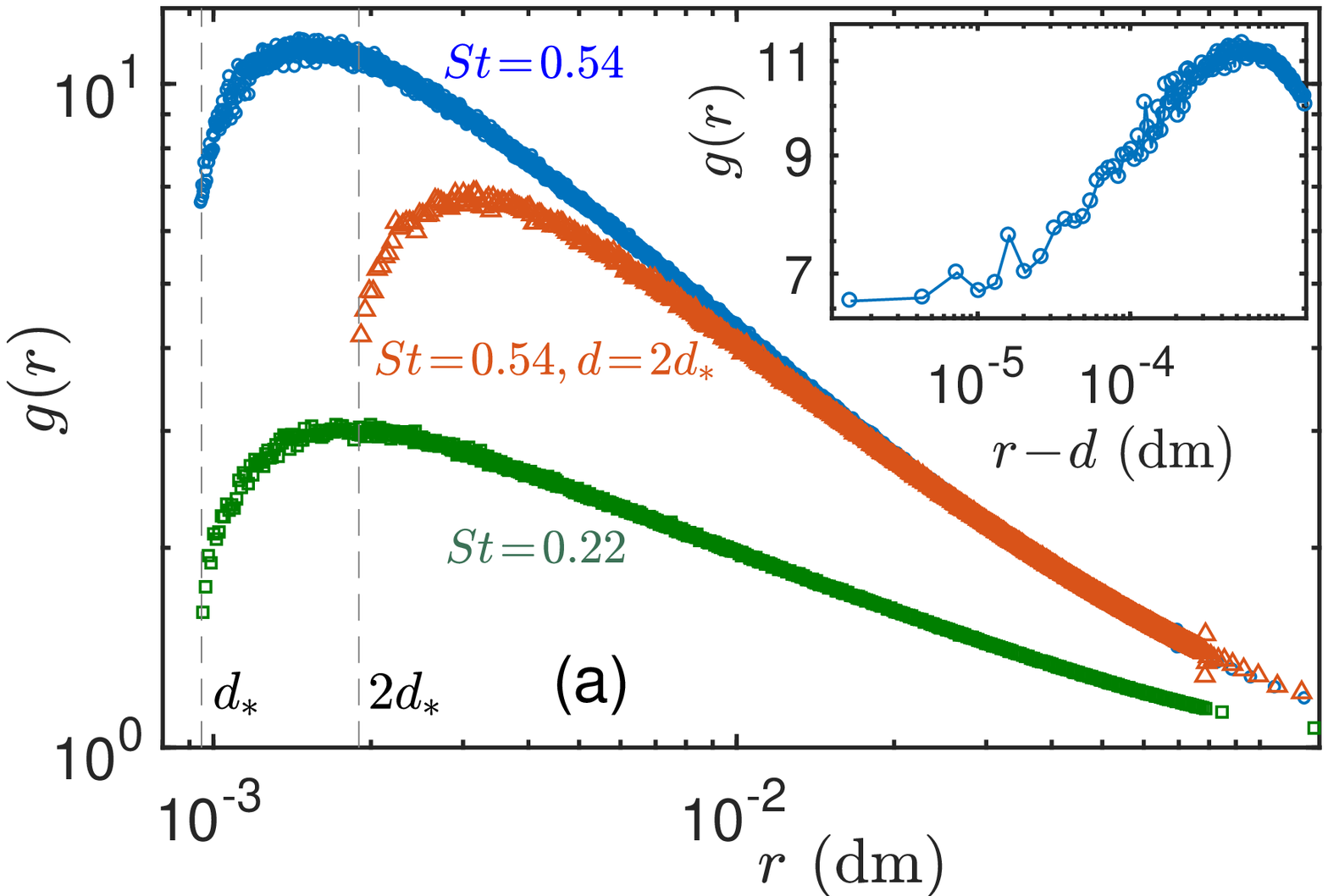}} \\ %\hspace{-20pt} 
     	\vspace{10pt}
     {\includegraphics[width=.49 \textwidth]{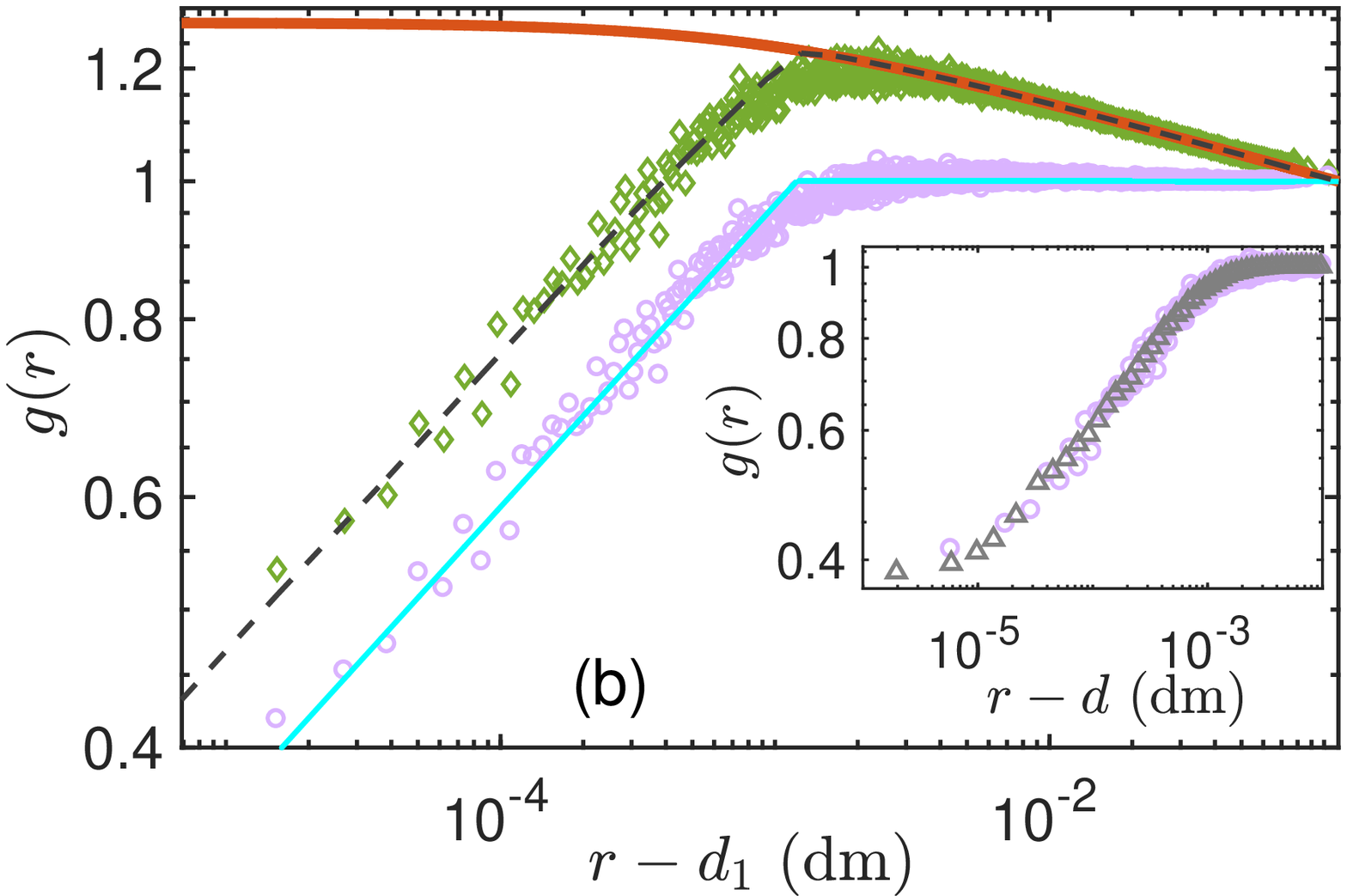}} %\hspace{-20pt}  \\ 
     \begin{picture}(0,0)
%		\put(-206,363){{\textcolor{blue}{$St\!=\!0.54$}}}
%		\put(-216,304){{\textcolor{Red}{$St\!=\!0.54, d\!=\!2d_*$}}}
%		\put(-206,264){{\color{OliveGreen}{$St\!=\!0.22$}}}
		%\put(-100,200){\includegraphics[height=1.8cm]{Figures/RDF1.eps}}
	\end{picture}
  \end{center}
  
\caption{\label{RDF1} \label{RDF2} \ew{RDFs $(\,g(r)\,)$ of particles that coagulate upon collision. {\bf a)} $g(r)$ for various cases of Stokes numbers and particle diameters ($d$). $\,\square\,$: $St\!=\!0.22$, $d\!=\!d_*$, $\,\Circle\,$: $St\!=\!0.54$, $d\!=\!d_*$, $\triangle\,$:~$St\!=\!0.54 $, $d\!=\!2d_*$. All $g(r)$ drop-off exponentially when $r \to d$ (details in text).
{\bf Inset:} $g(r)$ versus $r-d$ for the $\Circle$ case. It exemplify the fact that $\lim_{r \to d}\, g(r)$ is nonzero.
	{\bf b)} RDFs versus $r-d_1$ (where $d_1 \!=\! 0.99d$) for the case of $St\!=\!0.054$, $d\!=\!d_*$. $\Diamond \,$: the raw DNS-produced RDF $(g_{\text{DNS}}(r))$. Red-line: power-law fit to $g_{\text{DNS}}(r)$ (i.e., the $\Diamond$-plot) in the large-$r$ regime (the fit result is $0.890r^{-0.0535}$). %(fit-window: $r \!\in\! [0.6\eta,3\eta]\,$) 
It is equivalent to $g_s(r)$ in the ansatz $g_a(r)=g_0(r)g_s(r)$, i.e., it is the expected RDF for non-colliding particles under the same conditions. $\, \Circle \,$: the compensated RDF, i.e., $g_{\text{DNS}}(r)/g_s(r)$ (note: $g_s(r)$ is the red-line described earlier), this essentially gives us $g_0(r)$, which may be understood as a `modulation' on the RDF due to collision-coagulation. Cyan-line: two-piece power-law fits to the compensated RDF (the $\Circle$-plot) in the small and large $r-d_1$ regimes respectively (fit results: $4.17(r-d_1)^{0.212}$, $1.00(r-d_1)^{-2 \times\! 10^{-\!4}}$), this is an estimate for $g_0(r)$. Black-dashed-line: $g_0(r) g_s(r)$, (cyan-line $\times$ red-line), this shows that the ansatz accurately reproduces $g_{\text{DNS}}(r)$. 
	{\bf Inset:}  RDFs versus $r-d$. $\Circle \,$: compensated $g(r)$ for $St\!=\!0.054$, $d\!=\!d_*$, equivalent to the $\Circle$-plot in the panel's main figure; $\, \triangle \,$: compensated RDF for case $St\!=\!0.001$, $d\!=\!d_*$, i.e., finite size, almost zero $St$ particles  (in this case, the compensated and raw RDFs are the identical). This inset suggests that $g_0(r)$ has negligible St-dependence.}%\ew{can remove "center", add legends on the right and  save space?}
	}
\end{figure}

%\subsection{Results, theory and discussion}
We compute the RDF via $g(r) \!=\! N_{pp}(r)/[\frac{1}{2}N(N-1)\delta V_r/V]$, where $N_{pp}(r)$ is the number of particle pairs found to be separated by distance $r$, $\delta V_r$ is the volume of a spherical shell of radius $r$ and infinitesimal thickness $\delta r$, %$n_p$ is the global averaged particle number density.$V$ is defined earlier. %is the spatial volume of the domain.

Figure~\ref{RDF1} shows the \ew{RDFs obtained for monodisperse particles of various Stokes numbers and sizes. Two cases ($St\!=\!0.22$ and $0.54$) are shown in panel-a and two more ($St=0.054$ and $0.001$) are shown in panel-b. In this work, we focus on the smaller values of $St$ since
%this is where turbulence is expected to play a significant role in the collision-coagulation process in clouds before gravitational settling becomes dominant; 
the theory which we shall consider is also only applicable in the $St \ll 1$ regime. However, we have included the $St=0.54$ case to demonstrate that the observations to be described extends also to finite $St$.} In all cases, except one, the particles are of the same size $d=d_*$, where $d_*$ represents the specific value of $d_*\!=\!9.49\!\times\! 10^{-4}\, \text{dm}$, chosen so that the particle sizes are about $O(0.1)$ times the Kolmogorov scale ($\eta$), thus allowing us to still observe a regime ($3d \lesssim r \lesssim 30\eta$) of power-law RDFs. To shows the effect of changing particle size, panel-a also includes a case of $St\!=\!0.54, d\!=\!2d_*$ for comparison. \ew{Looking at panel-a, apart from the apparent  power-law behavior of the RDFs at intermediate values of $r$, the most striking feature of these RDFs for colliding-coagulating particles is that they fall-off dramatically in the $r \sim d$ regime. This is very different from what was seen in earlier studies of non-colliding particles where $g(r)$ are simple power-laws \citep{Chun05, Saw08}.} %(which increase monotonically as $r \to 0$). %The fall-off is so dramatic\footnote{We note that the abscissa is logarithmic, thus the RDFs are falling with decreasing $r$ with exponentially increasing steepness.} that it is unclear from simple inspection if the RDFs are still finite at particle contact ($r=d$) 
We also see that as $r$ approaches $d$ the steepness of the curve (see, e.g., the blue-circles) increases as $g(r)$ drops-off, this and the fact that the abscissa is logarithmic implies that $\frac{\partial g}{\partial r}$ is increasing exponentially in the process. As a consequence, it is difficult to discern from these plots if the limit of $g(r)$ at particle contact ($r \to d$) is still nonzero. This is an important question as $\lim_{r \to d}\left[g(r)\right]\!=\!0$ implies that the mean-field formula of \cite{Sundaram1997} has zero contribution towards $R_c$; i.e., collision rate is solely due to turbulent-fluctuations. It is only by re-plotting $g(r)$ versus $r\!-\!d\,\,$ (see insets in Fig.~\ref{RDF1}), and using a remarkable resolution that is $10^3$ finer than $d$, that we see a convincing trend supporting a finite $g(r \to d)$. %(This implies that the mean-field formula of \cite{Sundaram1997} still contributes towards $R_c$; otherwise, a non-zero $R_c$ is possible only via fluctuations.) %, i.e. $\left<w_i\,g(r, t)\right> \ne 0$.). 
 \ew{Also clear in panel-a is the observation that with changing particle size ($d$) the location of the sharp fall-off merely shifts to where the new value of $d$ is.}
 
The strong effect of particle collision on the RDF (also on MRV as we shall see later) challenges the validity of the "separation paradigm". We note that similar fall-off of RDF was previously observed \citep{Sundaram1997} but a complete analysis \ew{and theoretical understanding were lacking. %of its relation to other quantities and its implications was lacking;
Also,} a study on multiple collisions \citep{Vosskuhle13} had hinted at the potential problem with the separation paradigm.

Another observation is that in the power-law regime ($3d \lesssim r \lesssim 30\eta$), the RDFs appear (as expected) as straight-lines with slopes (i.e., power-law exponents) that increase with $St$ and are numerically consistent with those found for non-colliding particles (see, e.g., \citep{Saw12b}). 

\subsection{Theoretical Account via Drift-Diffusion Theory}

%To theoretically predict the new RDF, 
To theoretically account for the new findings, we make some derivations that are partially similar to the ones in \citep{Chun05}, but under a new constraint due to coagulations: At contact ($r=d$), the radial component of the particle relative velocities can not be positive\footnote{In other words particles may approach each other (and collide) but they can not be created at contact and then separate.}, %positive particle radial velocities ($w_r$) is not allowed at contact ($r=d$)%\footnote{Perhaps the colloquial term "no-escape boundary condition" may be elucidating here, in the same spirit as the well known "no-slip boundary condition". }
while with increasing $r$ the constraint is gradually relaxed. The first consequence of this is that the distribution of the radial component of the relative particle velocity ($W_r$) is highly asymmetric at $r \approx d$, i.e., the PDF of positive $W_r$'s are very small (this constitutes the "enhanced asymmetry" mentioned earlier). Thus for $r \approx d$, \ews{$\left<W_r\right>$ must be negative}. %(Note: this will lead to a nonzero mean-field contribution to $R_c$). %which is consistent with \cite{Sundaram1997}.) 
%From this, one could derive, in the limit of $St \ll 1$, a master equation (details in Sec.~\ref{derive_masterEqn} or \citep{Chun05}): %\cite{Supp}): 
In Sec.~\ref{derive_masterEqn}, we showed that in the $St \ll 1$ limit, one could derive a master equation (Eq.~\ref{master1}), reproduced here for clarity:
%\begin{equation} \label{master1}
\[ \frac{\partial \left< P \right>}{\partial t} + \frac{\partial}{\partial r_i} \left( q_i^d +q_i^D \right) +\frac{\partial (\left< W_i \right>\left< P \right>)}{\partial r_i} =0 \,, \]
%\end{equation}
%where $P(\bold{r})$ is the PDF of finding another particle at \ew{position} $\bold{r}$ from a `primary' particle\footnote{ borrowing the notation of CK-theory \citep{Chun05}, $W_i,P$ are ensemble-averages over trajectories of satellite (secondary) particles around a primary particle whose history (including the fluid's velocity gradient tensor around it) is known and fixed.}, $\left< . \right>$ implies averaging over all primary particle trajectories (e.g. $\left< W_r \right> \equiv$ unconditional mean of $w_r$), 
\ew{where $q_i^d$ is the drift flux (of probability due to turbulent fluctuation) defined in (\ref{def_q_d}) and $q_i^D$ is the diffusive flux defined in (\ref{def_q_D}).}

We then expand $W_i$, $\frac{ \partial W_l}{\partial r_l} $ and (consequentially) the fluxes as perturbation series with $St$ as the small parameter (details in \citep{Supp} or \citep{Chun05}). The coagulation constraint \ew{affects the values of the coefficients of these series. For the drift flux, the leading order terms (in powers of $St$) are:} %%\begin{widetext} 
\begin{equation} \label{drift_flux}
  q_i^d = -\left<P\right>\!(\mathbf{r})\,r_k \int_{-\infty}^{t} \left[ a^{(1)}_{ik} St\,  + a^{(2)}_{ki} St^2 \right]dt' \,,
 \end{equation} 
%\end{widetext}  
with \ew{ $a^{(1)}_{ik} \!=\! \tau_{\eta} \left< \Gamma_{ik}(t)\Gamma_{lm}(t')\Gamma_{ml}(t') \right>$ and
$a^{(2)}_{ki} \!=\! \tau_{\eta}^2 \left< \Gamma_{ij}(t)\Gamma_{jk}(t)\Gamma_{lm}(t')\Gamma_{ml}(t') \right>$,  $\Gamma_{ij}$ is the $ij$-th component of the fluid's velocity gradient tensor at the particle position (the $a^{}_{ik}$'s are thus related to two-time correlations of moments of velocity gradients, \cite{Chun05} shown that $a^{(2)}_{ik} \propto \overline{S^2} -\overline{R^2}$, where $(\overline{S^2},\,\overline{R^2})$ are the average fluid (strain rate tensor, rotation rate tensor) squared at particle positions)}. \ew{As explained earlier, coagulation-constraint causes the PDF of relative particle velocities to become highly asymmetric for $r \sim d$, thus $a^{(1)}_{ik}$ is nonzero at this scale. This is very different from the case of non-colliding particles \citep{Chun05} where $a^{(1)}_{ik}$ is always zero due to statistical isotropy.} Under the constraint, DNS gives $\int_{-\infty}^{t} a_{ik}^{(1)} dt' \approx \ew{-0.18\, s^{-\!1}}$	%$d_* \!\times\! \int_{-\infty}^{t} a_{ik}^{(1)} dt' \approx -0.2\!\times\!10^{-3}$ 
and $\int_{-\infty}^{t} a_{ki}^{(2)} dt' \approx \ew{2.45\, s^{-\!1}}$	%$ d_* \!\times\! \int_{-\infty}^{t} a_{ki}^{(2)} dt' \approx 2.9\!\times\!10^{-3}$
(more in \citep{Supp}). Thus \ew{for $r \sim d$, the drift flux is negative for large $St$ but becomes positive\footnote{Here a positive $q_r^d$ merely reflects a deficit in the inward flux of neighboring particles since we find that $q_r^d + q_r^D$ is always negative.}
when $St$ is below the value of $\approx 0.07$; and in the limit of $St \!\rightarrow\! 0$, it is dominated by the first term in (\ref{drift_flux})}.

$q_i^D$ is a `nonlocal' diffusion caused by fluctuations and can be estimated using a model that assumes the particle relative motions are due to a series of random uniaxial straining flows \citep{Chun05}. \ew{\cite{Chun05} showed that, generally,} $q_i^D$ has an integral form (due to nonlocality), and only in the special case where $g(r)$ is a power-law, may it be cast into a differential form (similar to a local diffusion). In view of the nontrivial $g(r)$ observed here, we must proceed with the integral form:
%\vspace{6pt}
%\scalebox{1}{% 
%\hspace{-10pt} $
\begin{multline}  
q_r^D = c_{st}\, r \! \\ 
\ \ \times \int\!\! d\Omega \!\int_0^{\infty}\!\! dt_fF(t_f) \!\!\int_{d/r}^{\infty}\!\! d\mbox{\scriptsize $R_0$} \,{\mbox{\scriptsize ${R_0}$}}\!^{2} \left< P\right>\!{(r \mbox{\scriptsize $R_0$})}\, f_I(\mbox{\scriptsize $R_0$},\mu,t_f) \,, \nonumber %$
\end{multline}
%} 

\vspace{5pt}
\noindent where $R_0 \equiv r_0/r$ with $r_0$ as the initial separation distance of a particle pair before a straining event, $F$ the probability density function for the duration of each event, $f_I$ is determined by relative prevalence of extensional versus compressional strain events (more details in \citep{Supp} or \citep{Chun05}) and \ew{$\Omega$ is the solid angle for the axis of the straining flow.} Note that due to coagulation, the $R_0$-integration starts from $d/r$. We differ crucially from the CK theory via the introduction of the \ew{(positive) factor $c_{st}$, which could be shown to equal $|c_1|$, where $c_1$ is the power law exponent of the RDF the particles would have assuming they are non-colliding (details in \citep{Supp}). }

%which is positive, of order $\lesssim 1$ and may depend on $St$ (more in \citep{Supp}); .

 \ew{By definition, $g(r) \equiv \alpha \left<P \right>$. Periodic boundaries in our DNS imply that $\alpha = V$,} 
%By definition, $g(r) \equiv \! \alpha \! \left<P \right>$, where $\alpha \! \equiv \! V/(4\pi r^2 \delta r)\,$, $\delta r$ is an infinitesimal radial increment 
(more in \citep{Supp}). Using this and the fact that the problem has only radial ($\,r\,$) dependence, we rewrite (\ref{master1}) as: 
 \begin{equation} \label{master2} 
r^2 \frac{\partial g(r,t)}{\partial t} + \frac{\partial}{\partial r} \left[ r^2 \alpha \left( q_i^d + q_i^D \right) +r^2 \left< W_r \right>g(r,t)\right]  =0 \,,
\end{equation}
where the content inside $[.]$ gives the total flux. For a system in steady-state, the first term in (\ref{master2}) is zero, and upon integrating %over $r$ 
with limits $[d, r]$, we have:
\begin{multline} \label{eqn_CR} %\label{eqn_CR_met} 
c_{st}\,r^3 \!\! \int \!d\Omega\! \int_0^{\infty} \!\! dt_fF(t_f)\! \int_{d/r}^{\infty}\! dR_0 \,R_0^2\, g{(rR_0)}\, f_I(R_0,\mu,t_f) \\
 + \ \ g(r)\left[ r^2\left< W_r \right> -A_{\tau}r^3 \right] = -R_c^* \,,
\end{multline}
where we have identified the total flux at contact ($r=d$) as the negative of the (always positive)  normalized collision rate $R_c^* \equiv R_c / \left(4\pi [N(N-1)/2]/V \right)$,  
and comparing with (\ref{drift_flux}), we see that:  
\begin{equation} \label{define_A}
A_{\tau} \equiv St \!\!\! \int_{-\infty}^{t}\!\!\!  a^{(1)}_{ik} dt' \,\,+\,\,  St^2\!\!\! \int_{-\infty}^{t}\!\!\! a^{(2)}_{ki} dt' \,,
\end{equation}  
%\begin{multline} \label{eqn_CR} 
%c_{st}\,r^3 \!\! \int \!d\Omega\! \int_0^{\infty} \!\! dt_fF(t_f)\! \int_{d/r}^{\infty}\! dR_0 \,R_0^2\, g{(rR_0)}\, f_I(R_0,\mu,t_f) \\
% + \ \ g(r)\left[ r^2\left< W_r \right> -Ar^3 \right] = -R_c^* \,,
%\end{multline}
%where, 
%Under the coagulation-constraint, the DNS gives $\int_{-\infty}^{t} a_{ik}^{(1)} dt' \approx -0.18$	%$d_* \!\times\! \int_{-\infty}^{t} a_{ik}^{(1)} dt' \approx -0.2\!\times\!10^{-3}$ 
%and $\int_{-\infty}^{t} a_{ki}^{(2)} dt' \approx 2.45$	%$ d_* \!\times\! \int_{-\infty}^{t} a_{ki}^{(2)} dt' \approx 2.9\!\times\!10^{-3}$
%(more in \cite{Supp}). Thus in the $r \sim d$ regime, the drift flux is positive (negative) for $St$ less (larger) than a value of order 0.01; and in the limit of $St \!\rightarrow\! 0$, the first term in (\ref{define_A}) dominates. 
\ew{with the specific values of the $t'$-integrals already given above. For clarity, we reiterate that on the left side of Eq.~(\ref{eqn_CR}), we have the diffusive flux ($q_r^D$), mean-field flux ($r^2g(r)\left< W_r \right>$), drift flux ($q_r^d$), while on the right, the total flux is given in terms of the normalized collision rate ($R_c^*$). We note that this equation embodies the full relationship among RDF, MRV and collision rate.} 
%and the total flux at contact which is equivalent to the negative of the (always positive) normalized collision rate: $R_c^* \equiv R_c / \left(4\pi [N(N-1)/2]/V \right)$; $r$ is the radial distance between two particles; $c_{st}$ is a function of $St$ (always positive and of order $\leq 1$, (more in \cite{Supp})); definitions of $\Omega, t_f, F, R_0, f_I, \mu$ can be found in "Methods". The coefficient of the drift-flux $A_{\tau}$ is a perturbative series with $St$ as the small parameter \citep{Chun05}, to leading orders: $A_{\tau} \equiv \int_{-\infty}^{t} [ a^{(1)}_{ik} St\,  + a^{(2)}_{ki} St^2\, ]dt'$. 

\subsection{Ansatz and Accuracy of the Theory}
Simple analytical solutions to Eq.~(\ref{eqn_CR}) may be elusive due to its integral nature \ew{(a consequence of the non-local diffusive-flux)}. %\footnote{we hope analytical solution may be possible by more mathematically gifted readers, perhaps with suitable approximations or asymptotic.}. 
However, one could gain \ew{insights into %its implications 
 it and test its accuracy via numerical solutions. To that end,} we begin with a simple ansatz for $g(r)$, \ew{then we curve-fit the ansatz to the DNS-produced RDF $\left(g_{\text{DNS}}(r)\right)$. This enables us to, firstly, verify that the ansatz could accurately represent $g_{\text{DNS}}(r)$, and secondly, obtain a "calibrated" ansatz that is a numerically accurate representation of $g_{\text{DNS}}(r)$. We then show that Eq.~(\ref{eqn_CR}), supplied with the calibrated-ansatz, could numerically predict $\left< W_r \right>\!(r)$ (i.e., the MRV) that agrees well with the DNS-produced MRV. In short, we will show that given a "correct" $g(r)$, (\ref{eqn_CR}) produces the "correct" $\left< W_r \right>\!(r)$.} 
 
The ansatz has the form $g_a(r)=g_0(r)g_{s}(r)$, with $g_s(r)=c_0r^{-c_1}$, \ew{i.e., the RDF form for non-colliding particles \citep{Chun05} under the same conditions. 
%  and $g_0(r)=c_{00}(r-d_1)^{c_{10}}$. The former is just the RDF for the collision-less case \citep{Chun05}. 
As a first order analysis, we let $g_0$, which embodies the effects of collision, takes the simplest form that could still capture the main features of the RDFs seen in Fig.\ \ref{RDF1}. Specifically, we let $g_0(r)=c_{00}(r-d_1)^{c_{10}}$, where $c_{00}(r),c_{10}(r)$ are each piecewise constant quantities that switch from their small-$r$ to large-$r$ values at a crossover-scale $r_c$ (of the order of $d$); i.e., $g_0$ is a two-piece power-law of $r-d_1$. (Note that our earlier finding of $g(r \to d) >0$ implies that $d_1 < d$.)} 

\ew{From a given DNS-produced RDF $\left(g_{\text{DNS}}(r)\right)$, we first obtain a calibrated $g_s$ by fitting $c_0r^{-c_1}$ to $g_{\text{DNS}}(r)$ in the power-law regime $d \ll r \lesssim 10\eta$ (see the red-line in Fig.~\ref{RDF2}b). Next, we compute the DNS estimate of $g_0$ via $g_0^{\text{DNS}}=g_{\text{DNS}}(r)/g_s(r)$ which is essentially a compensated RDF %($g_{\text{DNS}}$ compensated by $g_s$) 
(see the $\Circle$-plot in Fig.~\ref{RDF2}b). To get a calibrated $g_0$, we then fit the general form of $g_0$ given above to $g_0^{\text{DNS}}$ 
%to get the specific calibrated values of $c_{00}$,  $c_{10}$ and $d_1$} 
(see the cyan-line in Fig.~\ref{RDF2}b; note that each time, two pieces of power-laws are fitted to one $g_0^{\text{DNS}}$, and $r_c$ results naturally from the intersection of the two). %Our earlier finding of $g(r \to d) >0$ implies that $d_1 < d$. We also find that $d_1$ has negligible $St$-dependence when $St$ is small (see e.g. inset of Fig.~\ref{RDF2}b). 
Fig.~\ref{RDF2}b shows the calibrated ansatz for the case of $St\!=\!0.054$ and verify its accuracy (the red-line is $g_s(r)$, the cyan-line is $g_0(r)$ and the dashed-black-line ($g_0 \cdot g_s$) accurately reproduces $g_{\text{DNS}}(r)\,$). The inset in Fig.~\ref{RDF2}b shows that $g^{\text{DNS}}_0(r)$ (i.e., $ g_{\text{DNS}}(r)/g_s(r)$) is roughly $St$-independent for $St \ll 1$.} 
\begin{figure}
  \begin{center}
     {\includegraphics[width=.48 \textwidth]{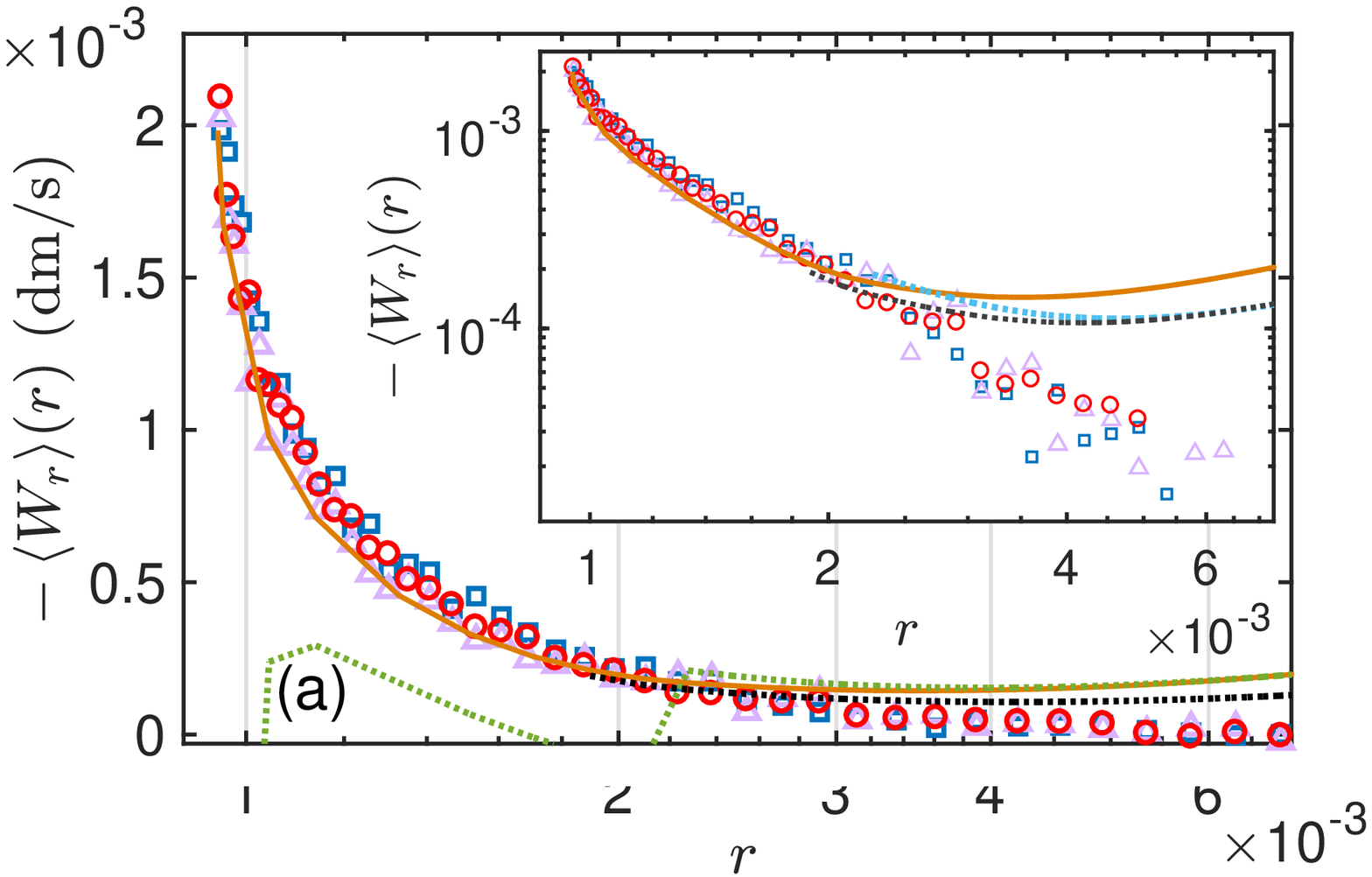}} \\ %\hspace{-20pt} 
     	\vspace{-20.9pt}
     {\includegraphics[width=.48 \textwidth]{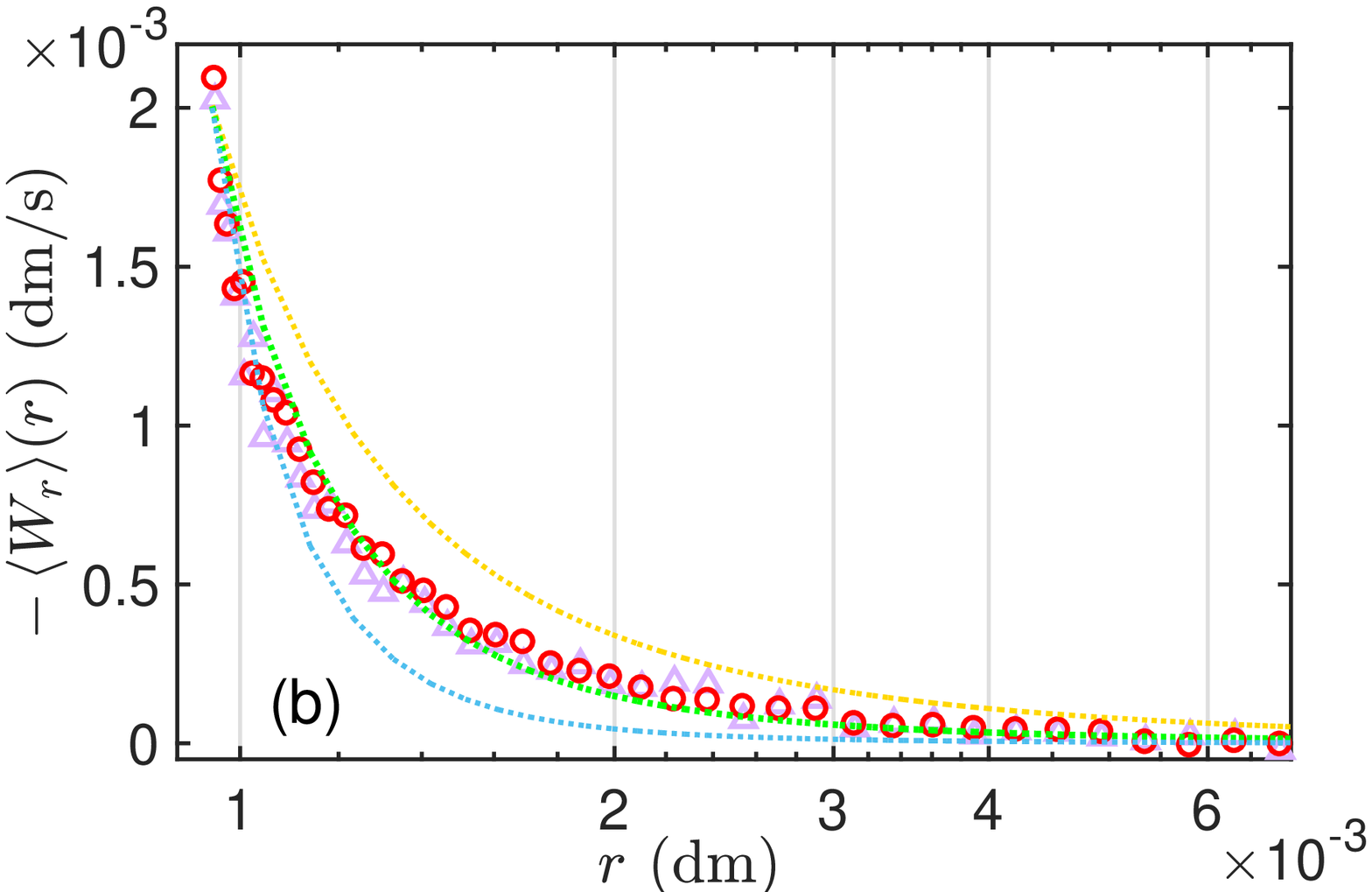}} %\hspace{-20pt}  \\ 
%     \begin{picture}(0,0)
%	\put(-80,80){\includegraphics[height=1.8cm]{Figures/RDF1.eps}}
%     \end{picture}
  \end{center}
\caption{\label{radial_vel} \label{radial_vel2}  Mean radial component of relative velocity (MRV) for particles \ew{of specific Stokes numbers and some theoretic-numerical predictions}. %Note: $(r,\,\left< W_r \right>)$ in units of (dm, dm/s). 
{\bf a)} Symbols are DNS results with $\,\triangle\,$: $St\!=\!0.001$; $\Circle\,$: $St\!=\!0.054$; $\,\square\,$: $St\!=\!0.11$. The lines \ew{are the numerical predictions by the theories (equation (\ref{eqn_CR}) or (\ref{eqn_CR_diff})) using the ansatz (details in text). Orange-line: $\left< W_r \right>^{\text{theory}}_{r \sim d, St=0.054}\,$}, i.e., the numerical prediction via the integral version of the theory (Eq.~(\ref{eqn_CR})) for the small-$r$ regime ($r \sim d$). Black-line: \ew{$\left< W_r \right>^{\text{theory}}_{r \gg d, St=0.054}\,$, same as the previous but for the large-$r$ regime ($r \gg d$). Green-line: prediction of the differential version of the theory (Eq.~(\ref{eqn_CR_diff})) for the $r \sim d$ regime. 
{\bf Inset)} A repeat of the main figure in log-log axes. Exception: Cyan-line is  the prediction of the differential version of the theory, but for the $r \gg d$ regime.}
	{\bf b)} \ew{MRV compared with predictions via the phenomenological model of particle approach angles (Eq.~(\ref{MRV_model}) and (\ref{Pangle})). DNS results: $\triangle$: $St\!=\!0.001$; $\,\Circle\,$: $St\!=\!0.054$. \ew{ Dotted lines are model predictions of} $\left< W_r \right>^{\text{}}_{St=0}$ using (\ref{MRV_model}) and (\ref{Pangle}) with variance $K$ obtained by matching the model's and DNS's transverse-to-longitudinal ratio of structure functions (TLR) of a certain order (from the top, yellow-line: order 2, green-line: order 4, cyan-line: order 6).}
}
\end{figure}

%For the purpose of demonstration and comparison, we take an approach consistent with \cite{Chun05} and consider the small $St$ limit ($St \ll 1$) which allows us to 
\ew{Next, we numerically evaluate the \ew{integral} in the first term of (\ref{eqn_CR}).
The $St \ll 1$ assumption allows us to approximate $g(r, St)$ inside the integral by its zero-$St$ cousin $g(r, St \to 0)$ \citep{Chun05}.
%As we are working in the small $St$ limit, we approximate $g(r, St)$ inside the integral using $g(r, St \to 0)$ (a practice borrowed from \citep{Chun05}). 
In practice, we replace $g(r, St)$ with the ansatz fitted to the DNS result of $g(r, St \!=\! 0.001)$. Next, we use the DNS data to estimate $A_{\tau}$, compute $R_c^*$ and $c_{st}$ (for this case, DNS gives $R^*_c= 9.69\!\times\! 10^{\!-\!10} \text{dm}^3\text{/s}$; $c_{st} = |c_1|$ as mentioned earlier). Finally we use (\ref{eqn_CR}) to predict $\left< W_r \right>\!(r)$.}
%we may approximate $g(r, St)$ in the integral in (\ref{eqn_CR}), to the first order, using $g(r, St \!\to\! 0)$ \citep{Chun05}. We thus numerically integrate the first term in (\ref{eqn_CR}) using the ansatz fitted to $g(r, St \!=\! 0.001)$; using the DNS data, we estimate $A_{\tau}$, obtain $R_c^*$ and $c_{st}$ (it can be shown that $c_{st} = |c_1|$ (more in \cite{Supp})), and \ew{then} use (\ref{eqn_CR}) to predict $\left< W_r \right>\!(r)$. 

\ew{Comparison of the predicted $\left< W_r \right>\!(r)$ with the ones obtained directly from the DNS is shown in Fig.\ \ref{radial_vel}. The prediction shown was made for the case of $St=0.054$, to be compared with its DNS counterpart (the $\Circle$ symbols). (We also show the DNS result for $St=0.001$ and $0.11$ to highlight an observation that $\left< W_r \right>\!(r)$ is almost $St$-independent in this small-$St$ regime.)} %As shown earlier, for $r \sim d$, $A_{\tau}$ is negative (positive) if $St$ is less (larger) than a value of order $0.01$. 
We have shown earlier that for $r \sim d$, $A_{\tau}$ is given by (\ref{define_A}). However, as stated earlier, as $r$ increases, the (statistical) asymmetry induced by collision-coagulation gradually becomes subdominant to the isotropy of turbulent-fluctuation. Statistical isotropy implies $a_{ik}^{(1)}=0\ $ \citep{Chun05}, a fact our DNS data confirm. Thus, for $r \gg d$, $A_{\tau}$ equals the order $St^2$ term in Eq.~(\ref{define_A}), exactly the same as the results of \citep{Chun05} for non-colliding particles.
%However, at larger $r$, the asymmetry induced by coagulation is gradually washed out by local isotropy of turbulence and $A_{\tau}$ relaxes to its \ew{collision-less} values, i.e.  $\lim\limits_{r \gg d} A = St^2\int_{-\infty}^{t} a_{ki}^{(2)}\!(t,t')\,dt' \approx St^2 a_{ki}^{(2)}\!(t,t)\,\tau_{\eta} $ \citep{Supp}. 
For this reason, we show two versions of \ew{the prediction: $\left< W_r \right>^{\text{theory}}_{r \sim d}$ and $\left< W_r \right>^{\text{theory}}_{r \gg d}$}, which are respectively obtained by setting $A_{\tau}$ to its small-$r$ and large-$r$ limits \ew{($-2.6\times\! 10^{-\!3}s^{-\!1}$, $\,7.1\times\! 10^{-\!3}s^{-\!1}$) respectively. The agreement between DNS and the predictions is noteworthy}, especially for small $r$. At $r \approx 2d$, the DNS result shows a weak tendency to first follow the upward trend of $\left< W_r \right>^{\text{theory}}_{r \sim d}$  and then drops off significantly \ew{at $r \gtrsim 2.5d $. The latter is consistent with the fact that $\left< W_r \right>^{\text{theory}}_{r \gg d}$ is below $\left< W_r \right>^{\text{theory}}_{r \sim d}\,$, but the drop is sharper than predicted.}

\subsection{Phenomenological Model of MRV}

Alternatively, \ew{(\ref{eqn_CR}) may be solved for the correct g(r), if $\left< W_r \right>$ is given.} \ew{As we are assuming $St \ll 1$, particle velocity statistics may be approximated by their fluid counterparts \citep{Chun05}, i.e., we may replace $\left< W_r \right>\!$ with $\left< W_r \right>\!_{St=0}$, the latter being the MRV of fluid particles. Hence, if $\left< W_r \right>\!_{St=0}$ is known, it may be used, together with (\ref{eqn_CR}), 
%one may assumes that in the small $St$ limit, particle velocity statistics are tied to their fluid counterparts \citep{Chun05}, thus (\ref{eqn_CR}) may be used, with $\left< W_r \right>\!_{St=0}$, 
to predict RDF of any finite but small $St$}. Fig.\ \ref{radial_vel}a shows that $\left< W_r \right>_{St>0}$ from the DNS do not change significantly for $St \in [0.001, 0.1]$, supporting this \ew{approach\footnote{This is true in the relatively idealized system simulated, but may not apply to the general problem that includes other effects}.}
 %In order to have a closed equation for $g(r)$, one must know the functional form of $\left< W_r \right>\!(r)$. As detailed in \cite{Chun05}, for $St \ll 1$, one may assume that all particle velocity statistics are given by their fluid-particle's ($St=0$) counterparts. 

Here we provide a simple, first order, model for $\left< W_r \right>\!_{St=0}$. We limit ourselves to the regime of small particles ($d \ll \eta$) and anticipate that $\left< W_r \right>$ is non-trivial (nonzero) only for $r \sim d$\ew{, a fact observable in Fig.~\ref{radial_vel}a}. We also assume that the relative trajectories of particles are rectilinear at such small scales. \ew{The coagulation constraint then implies} that: in the rest frame of a particle \ew{(call it P1), a second particle nearby must move in such a way that the angle ($\theta$) between its relative velocity and relative position (seen by P1) %which we denote as $\theta \in [0,\pi]$, 
must  satisfy: $\sin\!^{\!-\!1}(d/r) \,\le \,\theta\, \le \,\pi\,$, under the convention of $\sin^{\!-1}(x) \in [-\frac{\pi}{2},\frac{\pi}{2}]$,
%satisfy either $\theta > \frac{\pi}{2}$ or $\sin(\theta) \ge d/r$,
(more in \citep{Supp}). We can thus write (by treating negative and positive $w_r$ separately, applying the K41 theory \citep{Kolmogorov1941} and the bounds on $\theta$, details in \citep{Supp}), for $St \ll 1$, that:} 
%\begin{eqnarray*}
%\left< W_r \right> \equiv \left< w_r \right>_{*}  = p_- \! \left< w_r \, | w_r < 0 \right>_{*}  \,+\, p_+ \! \left< w_r \, | w_r \geq 0 \right>_{*}  
%\end{eqnarray*}
\begin{multline}  \label{MRV_model} 
\left< W_r \right> \equiv \left< w_r \right>_{*}  = p_{\text{--}} \! \left< w_r \, | w_r < 0 \right>_{*}  \,+\, p_{\text{+}} \! \left< w_r \, | w_r \geq 0 \right>_{*}  \\ \\
\ \ \ \ \ \ \  \approx -p_{\text{--}}\xi_{\text{--}} r \ +\ p_{\text{+}}\xi_{\text{+}} r \left[ 1 + \frac{\int_{\theta_m}^0 P^+_{\theta}\!(\theta')\cos(\theta')d\theta'}{\int_{0}^{\frac{\pi}{2}} P^+_{\theta}\!(\theta')\cos(\theta')d\theta'}\right] \,,
\end{multline}
where $\left< .\right>_{\!*}$ denotes averaging over particle pairs, $p_+$ ($p_-$) is the probability of a realization of $w_r$ being positive (negative), and $P^+_{\theta}$ is a conditional PDF such that $P^+_{\theta} \equiv P\left( \theta \  | \, w_r \ge 0 \right) \equiv P \left(\theta \  | \, \theta \in [0, \frac{\pi}{2}]\right)$, \ew{$\theta_m$ is the lower bound of $\theta$ described above}. % for $\theta$ conditioned on $w_r \ge 0$, or equivalently on $\theta \in [0, \frac{\pi}{2}]$.
For a first order account, we neglect skewness in the distribution of particle relative velocities and set $p_{\pm} =0.5$. \ew{Following \cite{Kolmogorov1941}, we have set $\left< w_r \, | w_r < 0 \right>_{\!*} = \xi_{\text{--}} r$, where $ \xi_{\pm} =C_s\sqrt{\varepsilon/(15\nu)}\,$, ($C_s$ is a Kolmogorov constant, we found $C_s=0.76$ by matching $\xi_{\text{--}}\, r$ to the first-order fluid velocity structure-function from the DNS)}. 

A simple phenomenological model for $P(\theta)$ may be constructed using the (statistical) central-limit-theorem by assuming that the angle of approach $\theta$ at any time is the sum of many \ew{random-incremental rotations in the past, thus we write}:
\begin{equation} \label{Pangle} 
P(\theta) = \it{N}\exp[K\cos(\theta-\mu_{\theta})] \sin(\theta) \,,
\end{equation} 
where $N\!\exp[...]$ is the circular normal distribution, i.e., analog of Gaussian distribution for angular data;  $\sin(\theta)$ \ew{results from} integration over azimuthal angles ($\phi$). We set $\mu_{\theta} = \frac{\pi}{2}$ (neglect skewness in fluid's relative velocity PDF) and obtain $K$ by matching the transverse to longitudinal ratio of structure functions (TLR) of the particle relative velocities with the ones via the DNS data; $N$ is determined via normalization of $P(\theta)$. Fig.\ \ref{radial_vel2}b shows the $\left< w_r \right>_{\!*}$ derived via (\ref{MRV_model}) and (\ref{Pangle}), using $K$ calibrated with TLR of 2nd, 4th, 6th order structure functions respectively. The results have correct qualitative trend of vanishing values at large $r$ that increases sharply as $r$ approach $d$, \ew{with the 4th-order's result giving the best agreement with DNS}. Currently we have not a satisfactory rationale to single out the 4th-order. The TLR of different orders give differing results may imply that our first-order model may be incomplete, possibly due to over-simplification in (\ref{Pangle}) or to the inaccuracy of the rectilinear assumption ($d/\eta$ in the DNS may be insufficiently small).

\subsection{Differential Version of the Theory, Its Validity and Solution}
\label{diff_version}
We now discuss an important but precarious theoretical issue. %that should be resolved in near future.
\ew{\cite{Chun05} clearly showed that the non-local diffusion ($q_r^D$) may be converted, from its general integral form, into a differential version only when the underlying RDF is a simple power-law. However, \cite{Lu10} and \cite{Yavuz18}, working in two very different scenarios, found that their predictions using the differential form of the theory agree well with experiments, even when the RDFs involved was clearly not power-laws. We shall attempt to remedy this apparent paradox in future work.} %If we take the same leap of faith and 
\ew{To examine how well this albeit unjustified method works here, we recast (\ref{eqn_CR}) into its differential form \citep{Chun05}:}   
 \begin{equation} \label{eqn_CR_diff} 
 -\tau_{\eta}^{-1}B_{nl}\,r^4\frac{\partial g}{\partial r} \,\,+\,\,  g(r)\left[ r^2\left< W_r \right> -A_{\tau}r^3 \right] = -R_c^* \,,
\end{equation}
where $B_{nl} =0.0397$ (this value is computed from our DNS, $B_{nl}$ is expected to depend on flow characteristics, e.g., $R_{\lambda}$ and $\tau_{\eta}$ (more in \citep{Supp}).
%($B_{nl}$ is expected to depend on $R_{\lambda}$, this value is computed using $\tau_{\eta}$ of our DNS (more in \cite{Supp}); in general, when RDF is not a power-law, $B_{nl}$ may also be $r$-dependent),
\ew{Using (\ref{eqn_CR_diff}), the same $g_sg_0$ ansatz, %and $A_{\tau}=7.1\times\! 10^{-\!3}\,s^{-\!1}$
 we make another prediction for $\left< W_r \right>_{St=0.054}$,  which is plotted in Fig.~\ref{radial_vel}a (dashed green line). 
This prediction is far from the DNS at $r \sim d$ but perform as well as the integral version at $r \gg d$ (the jump in the curve is just an artifact from the kink in the ansatz).} 
%The accuracy of the new prediction is worse (the jump correspond to the kink in the ansatz) but still on par with results above. 

\ew{One advantage of (\ref{eqn_CR_diff}) is that it allows for a general solution, which we now give, assuming $\left< W_r \right>$ is given} by (\ref{MRV_model}) \& (\ref{Pangle}):
\begin{equation} \label{gen_sol}
	g(r) = \frac{1}{\beta(r)} \left[ \int \beta(r) q(r) dr + C \right] \,,
\end{equation} 
with $q(r) = R_c^*\tau_{\eta}/(B_{nl}\, r^4) $, $\beta(r) = \exp\left[ \int p(r)dr \right]$ and $p(r) = \left[ \, A_{\tau}r - \left< w_r \right>_{\!*} \, \right]\tau_{\eta}/(B_{nl}\, r^2), $ (more in \citep{Supp}). 

\begin{figure}
  \begin{center}
     {\includegraphics[width=.46 \textwidth]{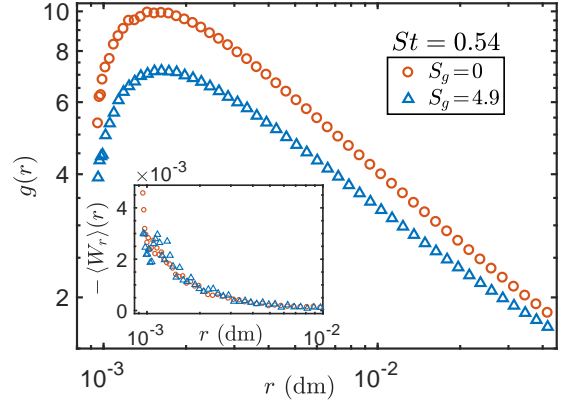}} \\ %\hspace{-20pt} 
     	%\vspace{-20.9pt}
     %{\includegraphics[width=.6 \textwidth]{Figures/Radial_Vel_2.eps}} %\hspace{-20pt}  \\ 
%     \begin{picture}(0,0)
%	\put(-80,80){\includegraphics[height=1.8cm]{Figures/RDF1.eps}}
%     \end{picture}
  \end{center}
\caption{\label{RDFnMRV_g1} RDFs of particles ($St=0.54$) subject to action of turbulence, collision-coagulation with and without gravity. Circles: $S_g=0$ (zero gravity); triangles: $S_g=4.9$ (nonzero gravity). The latter shows a reduced slope in the power-law regime, while the shape of the two curves largely similar in the collision regime ($r\sim d$). {\bf Inset)} MRVs of the same cases as in the main figure. Gravity weakens the MRV of the particles.}
\end{figure}

\ew{\subsection{Effects of Gravity and Other Limitations} \label{Limitations} \label{Gravity}
%\subsubsection{Effects of Gravity} \label{Gravity}
Thus far, we have not considered the effects of gravity on the particles. Here we provide a glimpse on the role of gravity (a detailed analysis is beyond the scope of the present study). In keeping with the scope of current work, we restrict ourselves to the case of monodisperse particle only. For this, we rerun the DNS cases of $St=0.054$ and $0.54$ with gravity (to be compared with the zero-gravity case). The new particle advection equation is: $d\mathbf{v}/dt = (\mathbf{u}-\mathbf{v})/\tau_p + \vec{g}\,$ (all other details of the DNS remain unchanged). We choose to have the particle settling parameter $S_g \equiv \tau_p\, g/u_{\eta}$ (where $u_{\eta}$ is the Kolmogorov velocity scale) be in the range $O(0.1)-O(1)$ %similar to the typical range of interest in the atmosphere or laboratory 
(this is achieved by letting $|\vec{g}|=10 \,\text{dm/s}^{2}$). \ews{As a result, the range of $S_g$ and $St$ explored here are well aligned with measured values in natural clouds \citep{Siebert10}.} For the case of $St=0.054$ ($S_g=0.49$), we find no discernible difference for both RDF and MRV between the "with gravity" and zero-gravity results (corresponding figures in \citep{Supp}). For the $St=0.54$ ($S_g=4.9$) case, Fig.~\ref{RDFnMRV_g1} shows the effects of gravity on the RDF and MRV. We see that the slope (exponent of $g(r)$ in the range $d \ll r < 20\eta\,$) of the RDF in the gravitational case is reduced by about 15\% compared to the zero-gravity case ($S_g=0$). %However, comparing the $g(r)$ of the $S_g=0$ case with the gravitational $g(r)$ multiplied by factor $1.4$ (the $\times$ plot), we see that the shape of the RDF is approximately preserved in the collision regime ($r\sim d$). 
However, the shape of the RDF in the collision regime ($r\sim d$) is approximately preserved, suggesting that a construct of the form $g_{\text{collision}} \times g_{\text{gravity}}$ may be a good first order model for the full RDF (close examination of the compensated RDFs gives substantial support for this idea, details in \citep{Supp}). These observations imply that as $S_g$ increase from $O(0.1)$ to $O(1)$, the effects of gravity on RDF grow from negligible to significant \ews{but not dominant}, the main effect is the reduction of the exponent while the collision related "modulation" ($g_{\text{collision}}$) remains largely intact. The inset of Fig.~\ref{RDFnMRV_g1} shows that the MRV is also weakened by gravity, albeit the statistical noise limits the strength of this conclusion. Lastly, It is worth noting that in the complimentary  DNS by \cite{Woittiez09} that included gravity but not actual collisions, much stronger gravitational effect was found on the statistics of bidisperse particles relative to the monodisperse case.}

%This implies that if practical applicability is of concern, the current results only applies to cloud particles with gravitational terminal velocities that are small compared to the velocity scale of the smallest turbulent eddies, e.g. particle of size $\lesssim 50 \mu m$ in atmospheric clouds). %(note also that the only viable theoretical treatment of this problem that we shall introduce in sequel assumes the asymptotic of very small particle which in turn implies limited importance of graviational setlling

\ew{As mentioned, the fundamental focus of our work precludes the DNS and theory from considering a number of complexities relevant to some applications. As a result, this limits the direct quantitative applicability of our results to some realistic problems (e.g., in clouds). Besides gravity, another neglected factor is the hydrodynamic inter-particle-force (HDI). Recent works, e.g., \cite{Yavuz18, Bragg22} found that HDI also has strong impact on RDF for $r \sim d$. For monodisperse particles with small to moderate $St$, HDI is expected to be more important than gravity. While we expect that HDI should not alter the qualitative trend that $g(r)$ should fall towards a small value at $r \to d$ (the same applies to the observed trend of MRV), it is likely that HDI and collision would affect RDF and MRV in a coupled manner.}

\ew{Also neglected is the influence of temperature, humidity and vapor-liquid phase transition which are important in the atmospheric clouds. These factors have substantial impact on the polydispersity of small droplets (see, e.g., \citep{Kumar12, Kumar14}). However, for monodisperse statistics considered here, they are likely to play minor roles (they will be more important when future works consider the full polydisperse problem). }

\ew{One limitation of the theory stems from the assumption of $St \ll 1$ and its corollary that particle velocity statistics in this regime are $St$-independent \citep{Chun05}, which limits the theory's applicability to real systems. This implies that MRV should be $St$-independent in this regime. Our DNS results (spanning two orders of magnitude in $St$) shown in Fig.~\ref{radial_vel} give some support to the latter. However, unlike the theoretical prediction for MRV of case $St=0.054$ (Fig.~\ref{radial_vel}), we have found that the prediction for $St=0.11$ is discernibly below the DNS result (figure in \citep{Supp}). This could be due to the finite $St$ effect not captured by the theory or other reasons (details in \citep{Supp}). Hence, a finite $St$ extension of the theory is desirable to improve its applicability to real systems.}

\conclusions  %% \conclusions[modified heading if necessary]

To conclude, we observed that collision strongly affects the RDF and MRV and imposes strong coupling between them\footnote{This statement also holds for other types of collisional outcomes (not only for collision-coagulation), but the details of the specific outcomes should be different from the current case.}. This challenges the efficacy of a "separation paradigm" and suggests that results from any studies that preclude particle collision has limited relevance for predicting collision statistics.
We have presented a theory for particle collision-coagulation in turbulence (based on a Fokker-Planck framework) that explains the above observations
%we took the CK theory \cite{Chun05} to uncharted regime of colliding and coagulating particles and verified that it is still accurate upon proper modifications (e.g. adding a symmetry-breaking constraint), i.e. we showed 
and verified its accuracy by showing that $\left< W_r \right>$ could be accurately predicted using a sufficiently accurate RDF. The theory accounts for the full collision-coagulation rate which includes contributions from mean-field and fluctuations, and as such, %this challenges the completeness of purely mean-field treatments 
our work complements and completes earlier mean-field theories \citep{Saffman1956, Sundaram1997}. We showed that a simple model of particle approach-angles could capture the main features of $\left< W_r \right>$ \ew{and applied it to derive a general solution for RDF from the differential version of the theory}.  \ew{We uncovered a possible paradox regarding the past empirical successes of the differential drift-diffusion equation (see Sec.~\ref{diff_version})}. %; we anticipate that a revised theory on the non-local diffusion will resolve this issue. 
\ew{Further shape-preserving reduction of the RDF and MRV were observed when gravitational settling parameter ($S_g$) is of order $O(1)$.}
Our findings provide new perspectives of particle collision and its relation with clustering and relative motion, which have implications for atmospheric clouds or generally for systems involving colliding particles in unsteady flows.
%any studies involving colliding particles in unsteady flows}
%We anticipate that an improved model for the particle approach-angles will be found and lead to higher order analytic solutions of (\ref{eqn_CR}) or (\ref{eqn_CR_diff}). Upon full appreciation of the consequences of particle collision, perhaps a treatment of the case of hard-sphere collision or inclusion of hydrodynamic forces is in order. 

%\subsection{HEADING}
%TEXT
%
%\subsubsection{HEADING}
%TEXT

%% The following commands are for the statements about the availability of data sets and/or software code corresponding to the manuscript.
%% It is strongly recommended to make use of these sections in case data sets and/or software code have been part of your research the article is based on.

%\codeavailability{TEXT} %% use this section when having only software code available
%
%
%\dataavailability{TEXT} %% use this section when having only data sets available
%
%
%\codedataavailability{TEXT} %% use this section when having data sets and software code available
%
%
%\sampleavailability{TEXT} %% use this section when having geoscientific samples available
%
%
%\videosupplement{TEXT} %% use this section when having video supplements available
%

\appendix
%\section{}    %% Appendix A
%
%\subsection{}     %% Appendix A1, A2, etc.

\noappendix       %% use this to mark the end of the appendix section. Otherwise the figures might be numbered incorrectly (e.g. 10 instead of 1).

%% Regarding figures and tables in appendices, the following two options are possible depending on your general handling of figures and tables in the manuscript environment:

%% Option 1: If you sorted all figures and tables into the sections of the text, please also sort the appendix figures and appendix tables into the respective appendix sections.
%% They will be correctly named automatically.

%% Option 2: If you put all figures after the reference list, please insert appendix tables and figures after the normal tables and figures.
%% To rename them correctly to A1, A2, etc., please add the following commands in front of them:

\appendixfigures  %% needs to be added in front of appendix figures

\appendixtables   %% needs to be added in front of appendix tables

%% Please add \clearpage between each table and/or figure. Further guidelines on figures and tables can be found below.

\authorcontribution{EWS oversaw the conception and execu- tion of the project. EWS did the theoretical derivations in collabo- ration with XM. XM and EWS conducted the numerical simulation and data analysis. EWS and XM wrote the article.} %% this section is mandatory

\competinginterests{The authors declared that none of them has any competing interests.} %% this section is mandatory even if you declare that no competing interests are present

%\disclaimer{TEXT} %% optional section

\begin{acknowledgements}
This work was supported by the National Natural Science Foundation of China (Grant 11872382) and by the Thousand Young Talent Program of China. We thank Jialei Song for helps. We thank Wai Chi Cheng, Jianhua Lv, Liubin Pan, Raymond A. Shaw for discussion and suggestions. 
\end{acknowledgements}

%% REFERENCES

%% The reference list is compiled as follows:

%\begin{thebibliography}{}
%
%\bibitem[AUTHOR(YEAR)]{LABEL1}
%REFERENCE 1
%
%\bibitem[AUTHOR(YEAR)]{LABEL2}
%REFERENCE 2
%
%\end{thebibliography}

%% Since the Copernicus LaTeX package includes the BibTeX style file copernicus.bst,
%% authors experienced with BibTeX only have to include the following two lines:

\bibliographystyle{copernicus}
\bibliography{biblio_all}

%%
%% URLs and DOIs can be entered in your BibTeX file as:
%%
%% URL = {http://www.xyz.org/~jones/idx_g.htm}
%% DOI = {10.5194/xyz}

\onecolumn
\section{Supplementary Material}

\subsection{Further Details of the Direct Numerical Simulation.}

\ew{The time step in our DNS $\triangle t$ is $0.001\,$s. The courant number is $C=0.073$, (where $C=\triangle t \left[\frac{u'}{\triangle x}+\frac{v'}{\triangle y}+\frac{w'}{\triangle z}\right]_{\text{max}}$, $u'$ etc. are r.m.s. velocities, $\Delta x$ etc. are grid spacings). The normalized maximum wavenumber simulated is $k_{\text{max}}\eta=1.2$. The turbulent flow is sustained by randomly forcing the two lowest nonzero shells of wave numbers. The integral length scale of the turbulent flow is estimated to be $L=0.646$~dm.} 

\ew{We study the statistics of monomers only (i.e., the particle of the same size ($d$) that we initially introduce into the system and which we later replenish at a constant rate close to the monomer-monomer collision rate). In this sense, the particle (monomers) are naturally lost from our consideration once they collide and become larger particles. Particles that become much larger ($St > 21.6$) are removed from the DNS at each time step.} 

\subsection{ Estimation of  Leading Order Terms in the Drift Flux, e.g $a^{(1)}_{ik}$}

Using the DNS data, we estimate, e.g., the value of 
\[ \int_{-\infty}^{t} a_{ik}^{(1)} dt'  \equiv \int_{-\infty}^{t} \tau_{\eta} \left< \Gamma_{ik}(t)\Gamma_{lm}(t')\Gamma_{ml}(t') \right> dt'. \] 
Note: the averaging is done over fluid particles (the theory assumed $St \ll 1$ limit, such that all velocity statistics are tied to the fluid's), the integrand is non-vanishing only for $t'$ in the vicinity of  $t- \tau_{\eta}$ to $t$ (where the turbulent velocity gradient $\Gamma_{ij}$ retains correlation), thus this quantity may be approximated as: $ {\tau_{\eta}}^2 \left< \Gamma_{ik}(t)\Gamma_{lm}(t)\Gamma_{ml}(t) \right> $.  As shown in \cite{Chun05}, $\left< \Gamma_{ik}(t)\Gamma_{lm}(t)\Gamma_{ml}(t) \right> $ is by definition zero in fully developed turbulence due to the fact that the small-scale statistics of turbulent flows are almost isotropic \cite{Kolmogorov1941}. However, the coagulation constraint dictates that at $r=d$, such averages must be taken with the condition that only fluid-particle pairs with negative radial velocity ($w_r < 0$) are taken into account (that the inertial particles' motion being tied to the fluid's does not imply that inertial pairs sample the fluid particle pairs's motion uniformly). Under this condition, the DNS data gives ${\tau_{\eta}}^2 \left< \Gamma_{ik}(t)\Gamma_{lm}(t)\Gamma_{ml}(t) \right>  \approx \ew{(-0.171}\times\!10^{-3} \,\text{dm/s})/d_*$, ($d_* =\! 9.49 \times\! 10^{-4} \text{dm}$); here, it is of value to point out that without such constraint or condition, the result for this quantity from the DNS is two orders of magnitude smaller. Similarly, we found $\int_{-\infty}^{t} a_{ki}^{(2)} dt' \approx \tau_{\eta}^3 \left< \Gamma_{ij}(t)\Gamma_{jk}(t)\Gamma_{lm}(t)\Gamma_{ml}(t) \right> \approx (\ew{2.32}\!\times\!10^{-3} \,\text{dm/s})/ d_*$; for this quantity, the DNS gives roughly the same values with or without the constraint.

\subsection{Full Definition of the Function $f_I(R_0, \mu, t_f)$ in the Model for Non-local Diffusive Flux.}

Derived in \cite{Chun05}, summarized here (with typo corrected), the diffusive action of the turbulence on the particle-pairs is assumed to consist of a random sequence of uniaxial extensional or compressional flows defined, and:
\[ f_I(R_0, \mu, t_f) \equiv f_+ I_+(R_0, \mu, t_f) + f_- I_-(R_0, \mu, t_f) \,, \]
where $R_0 \equiv r_0/r$, $r_0$ is the initial separation distance of a particle pair before a straining event, $r$ is the independent variable of the equation for $g(r)$; $f_+$ and $f_- \equiv 1 -f_+$ are the fractions of those flows that are extensional and compressional, respectively. \ew{Comparing with DNS, \cite{Chun05} calibrated $f_+$ and found $f_+ = 0.188$ (a result we use here)}. $I_{\pm}$ is an indicator function such that it takes the value $+1$  ($-1$) when a secondary particle leaves (enters) a sphere of radius $r$ centered on the primary particle, and otherwise zero. $\mu$ is the cosine of the angle between the axis of symmetry of the straining flow event and the \ew{displacement vector between} the two particles, $t_f$ is the lifetime of the event. To obtain a strain rate correlation function that decays exponentially with a characteristic time scale $\tau_S$, \cite{Chun05} set the probability density function for $t_f$ to be:
\[  F(t_f) = \frac{f_s t_f}{{\tau_S}^2} \exp(-t_f/\tau_S) \,. \]

The indicator function is used to count the net loss of particles from within the sphere over the duration of an (extensional or compressional) event and can be expressed as: 
\[  I_{\pm}(R_0, \mu, t_f)=H(1 -R_0)H(R_{f\pm} -1) \, - \, H(R_0 -1)H(1 -R_{f\pm}), \]
where $H(x)$ is the Heaviside function (zero for $x<0$, unity for $x \ge 0$), $R_{f\pm}$ is the non-dimensional final position of a particle pair with an initial position of $R_0$  and can be written as:
\ew{ \[  R_{f+} = R_0 \left[  \mu^2{\theta_t}^2 + \frac{(1-\mu^2)}{\theta_t} \right]^{1/2} \,, \]
\[  R_{f-} = R_0 \left[  \frac{\mu^2}{{\theta_t}^2} + {(1-\mu^2)}{\theta_t} \right]^{1/2} \,, \]
for uniaxial extension and compression respectively, where:
\[  \theta_t \equiv \exp \left( \frac{t_f}{\tau_{\eta} \sqrt{3 f_s}} \right) \,. \]
\newline }

\begin{figure}
  \begin{center}
 	{\includegraphics[width=.55 \textwidth]{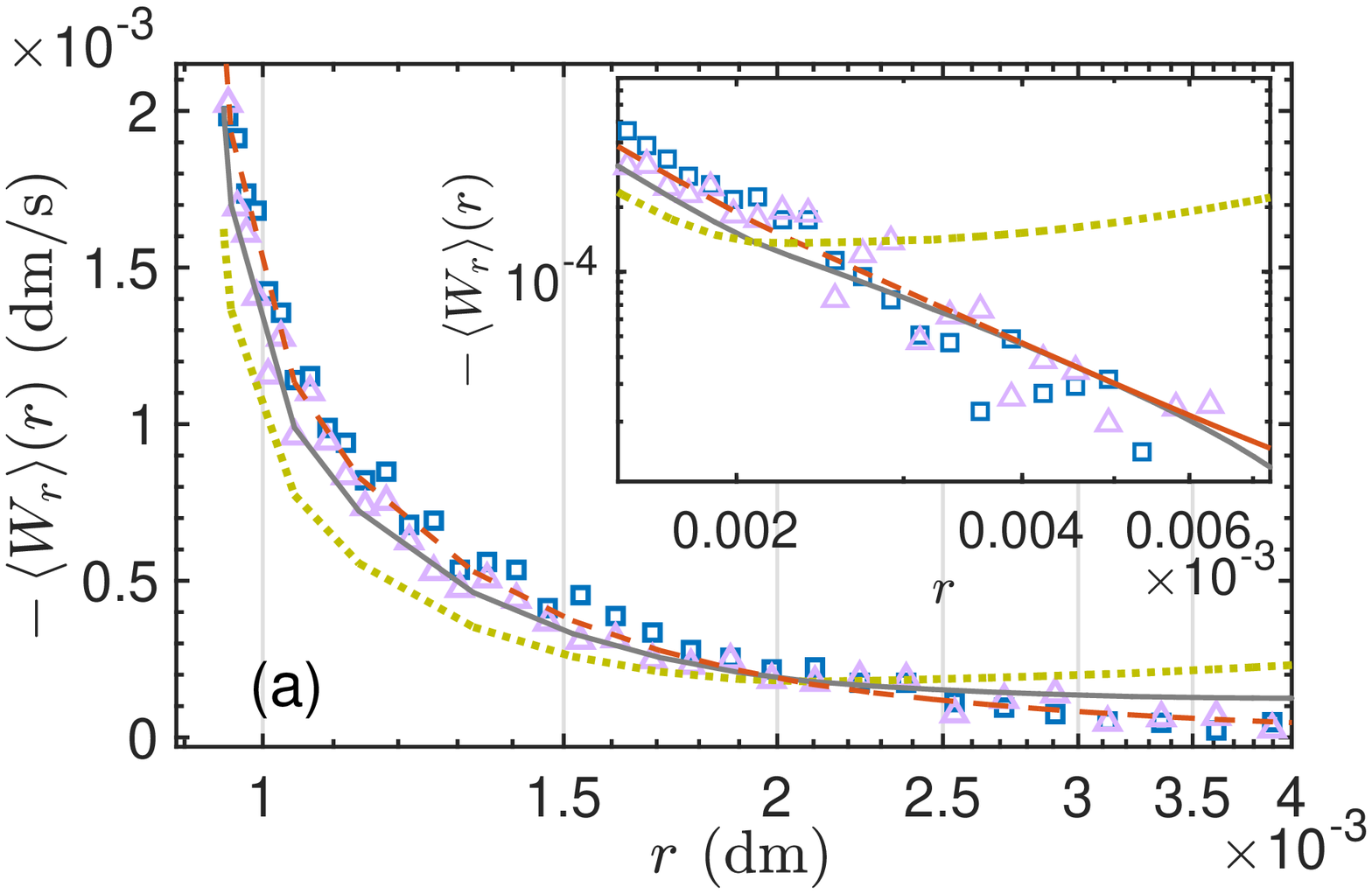}} \\ %\hspace{-20pt} 
     	%\vspace{-20.9pt}
%     \begin{picture}(0,0)
%	\put(-80,80){\includegraphics[height=1.8cm]{Figures/RDF1.eps}}
%     \end{picture}
  \end{center}
\caption{\label{radial_vel_more_St_supp} MRVs of particles. Triangles: DNS result for $St=0.001$. Squares: DNS result for $St=0.11$. Red-dashed-line: theory's prediction for $St=0.001$. Gold-dotted-line: theory's prediction for $St=0.11$. Grey-solid-line: modified-theory's prediction for $St=0.11$ (details of modification in \ews{the section on derivation of $c_{st}$}). Note: the predictions are based on the $A_{\tau}$ values valid in the $r \sim d$ regime ($d\!=\!9.49\!\times\! 10^{-4}$ \text{dm}). {\bf Inset)} Similar plots in logarithmic axes highlighting the large-$r$ regime. Note: the predictions are based on the $A_{\tau}$ values valid in the $r \gg d$ regime. It is clear that the modified-theory's predictions agree much better with the DNS.}
\end{figure}

\ew{\subsection{MRV Predictions by the Theory for Other Stokes Numbers.}
As mentioned in the main text of this manuscript, even though the theory assumes that MRVs are $St$-independent for small $St$'s. Nevertheless, it could produce separate predictions for each $St$. Here we show the predictions for $St=0.001$ and $0.11$ in Fig.~\ref{radial_vel_more_St_supp}. The prediction for $St=0.001$ (red dash line) agrees well with the DNS results (symbols), but the prediction for $St=0.11$ (gold dotted line) deviates significantly, suggesting that finite $St$ effects not captured by the theory start to become significant and thus diminish the accuracy of the theory.}

\subsection{\ew{Derivation of {\large $c_{st}$}, its Role and Possibility of Further Corrections to The CK Theory.}} \label{Further_correct}
%We begin by briefly noting that the model for non-local diffusion in the original CK theory is a relatively ad-hoc model with some multiplicative factors that may well be functions of Reynolds number. 
\ew{In this work, we deviate crucially\footnote{'Crucial' refers to the fact that without $c_{st}$ the theory would be inconsistent with previous experimental results (as this section will show)  and it would also produces results far from our DNS results.} from the CK theory \cite{Chun05} by introducing an extra %(correction)
factor $c_{st}$ (positive, of order unity or less) in the model of non-local diffusion:}
\begin{equation} \label{define_qD}
  q_r^D \!=\! c_{st}\, r \!\int d\Omega \int_0^{\infty}dt_fF(t_f) \int_{d/r}^{\infty} dR_0 \,{R_0}^2 \left< P\right>\!{(r \mbox{\scriptsize $R_0$})}\, f_I(\mbox{\scriptsize $R_0$},\mu,t_f) \, . 
\end{equation}
 To determine what $c_{st}$ is (or should be), we begin from an important finding in \cite{Chun05} that if $\left< P \right>$ is power-law of $r$, i.e., $\left< P \right> = C r^{-c_1}$, then the non-local diffusion $q_r^D$ can be cast into a differential form (which is usually only true for local diffusion):
\begin{equation} \label{define_qD_diff}
 q_r^D = -B_{nl}\, {\tau_{\eta}}^{\!-1}\, r^2 \,\frac{\partial \left< P \right>}{\partial r}  \,, \\
 \end{equation}
where:
\begin{equation} \label{define_Bnl}
B_{nl}  = \tau_{\eta} \int d\Omega \int_0^{\infty}dt_fF(t_f) \int_{d/r}^{\infty} dR_0 \,{R_0}^{2-c_1} f_I(\mbox{\scriptsize $R_0$},\mu,t_f) \, .
\end{equation} 
\vspace{5pt}
This, together with: $q_i^d = - A_{ck}\, {\tau_{\eta}}^{\!-1} \, r \left<P\right>$, eventually leads to the first order equation differential equation for the RDF \ew{($\,g(r)  \equiv V \left<P\right>\,$)}, that has (only) power-law solutions: $g(r) = VC r^{-c_1}$. % with: 
%\begin{equation} \label{c1_in_AnB}
% c_1 = \frac{A_{ck}}{B_{nl}} \]
%\end{equation}
This result \ew{(i.e., $g(r)$ or equivalently $\left< P \right>\!(r)$ are power-laws)} has seen compelling validations from both experiments (e.g., \cite{Saw12b, Lu10, Yavuz18}) and DNS (e.g., \cite{Chun05, Bec07, Saw12a}). We now begin from this experimentally validated result and work backward to derive an expression for $c_{st}$. We plug the power-law form for $\left< P \right>$ into (\ref{define_qD_diff}):
%\begin{widetext}
\begin{align*}
q_r^D &= -B_{nl}\, {\tau_{\eta}}^{\!-1}\, r^2 \,\frac{\partial (C r^{-c_1})}{\partial r}  \\
& = -B_{nl}\, {\tau_{\eta}}^{\!-1}\, r^2 C (-c_1) r^{-c_1-1} \\
& = B_{nl}\, {\tau_{\eta}}^{\!-1}\, r c_1\,C\, r^{-c_1} \\
& =  {\tau_{\eta}}^{\!-1}\, r c_1\,C\, r^{-c_1} \tau_{\eta} \int d\Omega \int_0^{\infty}dt_fF(t_f) \int_{d/r}^{\infty} dR_0 \,{R_0}^{2 -c_1} f_I(\mbox{\scriptsize $R_0$},\mu,t_f) \\
& = r c_1 \int d\Omega \int_0^{\infty}dt_fF(t_f) \int_{d/r}^{\infty} dR_0 \,{R_0}^{2}\, C(rR_0)^{-c_1} f_I(\mbox{\scriptsize $R_0$},\mu,t_f) \\
& = c_1\, r \int d\Omega \int_0^{\infty}dt_fF(t_f) \int_{d/r}^{\infty} dR_0 \,{R_0}^{2}\, \left< P\right>\!{(r \mbox{\scriptsize $R_0$})}\, f_I(\mbox{\scriptsize $R_0$},\mu,t_f) \,. \\
\end{align*}
%\end{widetext}
Comparing with (\ref{define_qD}), we have:
\[ \mbox{\large $c_{st} = |-c_1| \equiv |c_1|$} \,,\]
which is found in experiments (and theories) to be of order $0$ to $1$ and a function of particle Stokes number $St$; in words, this means $c_{st}$ is given by the modulus of the power-law exponent of the RDF that would arise in the collision-less case; in the case with collision and sufficiently small particle ($ d/\eta \lesssim 1\,$), such as in this study, $c_{st}$ equals the modulus of the power-law exponent of the RDF  the range of $d \ll r \ll 20\eta\,$ (note: power-laws RDF are empirically observed for $r \ll 20\eta\,$ \cite{Saw08, Saw12a}). Note: we have chosen to  define $c_{st }$ using the `modulus' (instead of the `negative' of the power-law exponent) since it guarantees that $q_r^D$ is negative (positive) when $g(r)$ is an increasing (decreasing) function of $r$, so that we are consistent with the fact that $q_r^D$ is a diffusion flux.
 We note that both the CK theory and the current modified version assume $St \ll 1$. %Finally we note that this new model for $q_r^D$ gives the correct asymptotic when $St \to 0$.
 
Chun et al. \cite{Chun05} went further to provide a solution for $c_1$ (for collision-less particles, in the $St \ll 1$ limit):
\begin{equation} \label{c1_in_AnB}
	\ew{ c_1 = \frac{A_{ck}}{B _{nl}} \equiv \frac{A_{\tau, r\! \gg \! d} \,\, \tau_{\eta}}{B_{nl}}\,, } 
\end{equation}
where \ew{we have clarified that $A_{\tau}$ in our work is defined differently from "$A$" in \citep{Chun05} (we denote the latter as $A_{ck}$ to avoid confusion), and $A_{\tau, r\! \gg \! d}$ is our $A_{\tau}$ evaluated at the large-$r$ limit.} In the current context, $c_1$ maybe obtained via (\ref{c1_in_AnB}) or alternatively directly from the power-law exponent of $g(r)$ 
in the range $d \ll r \ll 20\eta\,$ as discussed above. Using values of the relevant parameters in our DNS, we found $\frac{A_{\tau}\, \tau_{\eta}}{B_{nl}} \approx \frac{2.4 St^2 \times .0925}{.0397} = 5.6 St^2$, which is $15\%$ smaller than the one found in \cite{Chun05}, i.e., $ \frac{A_{ck}}{B_{nl}} \approx  \frac{.61 St^2}{.0926} = 6.6 St^2$. However, we have observed in our DNS that the direct method (by fitting power-laws to the RDFs in the suitable $r$-range) gives $c_1$ which is $3.2$ ($1.9$) times larger than the one obtained using (\ref{c1_in_AnB}) for the case of $St=0.054$ (0.11).

A plausible interpretation of the discrepancy described just above is that there may be another missing dimensionless factor (of order unity, possibly weakly dependent on Reynolds-number) in the correct definition of $q_r^D$ . \ew{This is beyond the scope of this present study (to avoid confusion, we currently restrict ourselves to the least speculative correction only)} and is a good subject for future works. However it may be informative to note that, by inspection, we find that if we further include a factor of $\sim 1/3\ \text{to}\ 1/2$ in the definition of $q_r^D$, then the agreement  between the theoretical (the integral version) and DNS produced $\left< W_r \right>$ is strikingly better in the $r \gg d$ limit, while in the $r \sim d$ regime, it is slightly better (the former should not come as a surprise as this is the regime of power-law RDFs and the factor of $\sim 1/3$ is exactly designed to reproduce the correct $c_1$). %For the differential version of the theory, we find that the improvement is decisively strong for all $r$.

\ew{To demonstrate the point just discussed, we show in Fig.~\ref{radial_vel_more_St_supp} the predictions by the theory for the case of $St=0.11$. We see that, in the $r \sim d$ regime, the prediction by the original theory (dotted line in the main figure) is somewhat below the DNS result, while the prediction by the modified theory (with a factor of $1/2$ appended to the definition of $q_r^D$), shown as the solid line, is much closer to DNS. In the $r \gg d$ regime, the modified theory's superiority in terms of accuracy is even more pronounced (see inset of Fig.~\ref{radial_vel_more_St_supp}).}

\subsection{Relation Between $g(r)$ and $\left< P \right>$.}

In the main text, we state that $g(r) \equiv V \left<P \right>$, where $V$ is the spatial volume of the full domain of the problem, i.e., $(2\pi)^3$ in the DNS. Justification: let $g(\vec{r})$ be the ratio of probability of finding a second particle at $\vec{r}$ from a particle, to the probability of such finding in a perfectly random distributed particle population, thus: $g(\vec{r}) \equiv \frac{\left< P \right> \delta x \delta y \delta z}{(\delta x \delta y \delta z)/V} \equiv \left< P \right> V$. Further, since system is isotropic, $g(\vec{r}) \equiv g(r)\,$.

%In the main text, we defined $\alpha$ via $g(r)=\alpha \left<P \right>$, where $\alpha \equiv V/(4\pi r^2 \delta r)$ , $\delta r$ is an infinitesimal radial increment. Note that $4\pi r^2 \delta r \equiv \delta V_r$ , i.e. the volume of a spherical shell of radius $r$ and infinitesimal thickness $\delta r$, which appear elsewhere in the paper;
%while $V$ is the spatial volume of the domain i.e. $(2\pi)^3$ in the DNS.  

\subsection{Modeling of MRV based on Distribution of Particle Approach Angles $P(\theta)$.}

We imagine the particles are small, i.e., $d \ll \eta$ and $St \ll 1$. The latter implies their trajectories are almost like fluid particles', while the former implies that, viewed at the scale of interest $r\sim d$, their trajectories are almost rectilinear (since the radii of curvature are proportional to $\eta$). Thus in the reference frame of a primary particles, no secondary particle could have a trajectory, being straight-line, that has a history of collision with the volume of the primary (otherwise coagulation would have occurred and the secondary particle in question would cease to exist).  In trigonometric terms, let $\theta$ be the angle between the secondary particle's velocity and its vector position in the rest frame of the primary particle, then we must have: $\sin\!^{\!-\!1}(d/r) \,\le \,\theta\, \le \,\pi\,$, with the convention that $\sin^{\!-1}(x) \in [-\frac{\pi}{2},\frac{\pi}{2}]$. 
%, this translates to $\pi \ge \theta \ge \sin^{-1}(d/r)$ with condition that $\sin^{-1}(x) \in [-\frac{\pi}{2},\frac{\pi}{2}]$, where $\theta$ is the angle between the secondary particle's velocity and its vector position in the frame of the primary's. 

From the above, we could then compute the MRV, $\left< w_r \right>_{\!*}$ \ew{based on fluid particles' statistics. Since collision-coagulation affects positive and negative relative particle velocities differently, we begin by writing $\left< w_r \right>_{\!*}$} as a sum of the positive (i.e., $w_r > 0$) and negative branches (with proper statistical weights $p_{\pm}$ to account for possible skewness of the probability distribution of velocity):
\[ \left< W_r \right> \equiv \left< w_r \right>_{*}  = p_- \! \left< w_r \, | w_r < 0 \right>_{*}  \,+\, p_+ \! \left< w_r \, | w_r \geq 0 \right>_{*}  \, . \] 
The negative branch  $p_- \! \left< w_r \, | w_r < 0 \right>_{*}$ is unaffected by collision-coagulation and we thus express it as a simple linear function of $r$ that follows from the K41-phenomenology \citep{Kolmogorov1941}, i.e., $-p_- \,\xi_- \, r$, where $\xi_{\pm} \sim \sqrt{\varepsilon/(15\nu)}$, $\varepsilon$ is the (kinetic) energy dissipation rate of the flow. For the positive branch, we further assume that the \ew{(fluid particles')} joint probability density function (PDF) of magnitude of relative velocity (secondary particle relative to primary particle) $|\vec{w}|$  and approach-angle $\theta$, $P(|\vec{w}|, \theta)$, is separable \ew{(note: $w_r \equiv |\vec{w}| \cos(\theta)$)}, hence:
%\begin{eqnarray*}
\begin{align*}
& p_+ \! \left< w_r \, | w_r \ge 0 \right>_{*}  \\
& \ \ = \int_0^{\infty} \!\! d|\vec{w}| \! \int_{\theta_m}^{\frac{\pi}{2}} \! d{\theta} \,P(|\vec{w}|, \theta) \, |\vec{w}| \cos(\theta)  \\
& \ \ = \int_0^{\infty} \!\! d|\vec{w}| \, P_w(|\vec{w}|) \,|\vec{w}| \int_{\theta_m}^{\frac{\pi}{2}} \! d{\theta} \,P_{\theta}(\theta) \, \cos(\theta) \\
& \ \ = p_+ \int_0^{\infty} \!\! d|\vec{w}| \, P_w(|\vec{w}|) \,|\vec{w}| \int_{\theta_m}^{\frac{\pi}{2}} \! d{\theta} \,P_{\theta}^+(\theta) \, \cos(\theta) \,,
\end{align*}
%\end{eqnarray*}
where all the $P$'s are PDFs, note that $p_+ \equiv \int_0^{\frac{\pi}{2}} P_{\theta} \,d\theta\,$, $\int_0^{\frac{\pi}{2}} P_{\theta}^+ d\theta \equiv \int_0^{\frac{\pi}{2}} \left( P_{\theta}/p_+ \right) d\theta =1$ and $\int_0^{\pi} P_{\theta} \, d\theta =1$, \ew{also note that $P_{\theta}^+ \equiv P_{\theta}(\,\theta \,\,\, | \,w_r \ge 0)$}; more importantly $\theta_m = \sin^{-1}(d/r)$ as previously explained.
Further:
%\begin{widetext}
\begin{align*}
& p_+ \! \left< w_r \, | w_r \ge 0 \right>_{*} 		\\
& \ \ = p_+ \int_0^{\infty} \!\! d|\vec{w}| \, P_w(|\vec{w}|) \,|\vec{w}| \int_{\theta_m}^{\frac{\pi}{2}} \! d{\theta} \,P_{\theta}^+(\theta) \, \cos(\theta) 	\\
& \ \ = p_+ \int_0^{\infty} \!\! d|\vec{w}| \, P_w(|\vec{w}|) \,|\vec{w}| \left[ \int_{0}^{\frac{\pi}{2}} \! d{\theta} \,P_{\theta}^+(\theta) \, \cos(\theta) +  \int_{\theta_m}^{0} \! d{\theta} \,P_{\theta}^+(\theta) \, \cos(\theta) \right]	\\
& \ \ = p_+ \int_0^{\infty} \!\! d|\vec{w}| \, P_w(|\vec{w}|) \,|\vec{w}|  \int_{0}^{\frac{\pi}{2}} \! d{\theta} \,P_{\theta}^+(\theta) \, \cos(\theta) \left[ 1 +  \frac{\int_{\theta_m}^{0} \! d{\theta} \,P_{\theta}^+(\theta) \, \cos(\theta)}{ \int_{0}^{\frac{\pi}{2}} \! d{\theta} \,P_{\theta}^+(\theta) \, \cos(\theta)} \right]	\\
& \ \ = p_+ \xi_+ \, r \left[ 1 +  \frac{\int_{\theta_m}^{0} \! d{\theta} \,P_{\theta}^+(\theta) \, \cos(\theta)}{ \int_{0}^{\frac{\pi}{2}} \! d{\theta} \,P_{\theta}^+(\theta) \, \cos(\theta)} \right] \,,
\end{align*}
%\end{widetext}
where in the last line, we have replaced the \ew{first two integrals, combined,} with its K41 estimate, where $\xi_{\pm} \sim \sqrt{\varepsilon/(15\nu)}$ .

\subsection{Prediction of the Peak Location of the RDF Using the Differential Form of the Drift-Diffusion Equation.}

 \begin{equation} \label{eqn_CR_diff_supp} 
 -\tau_{\eta}^{-1}B_{nl}\,r^4\frac{\partial g}{\partial r} + \ \ g(r)\left[ r^2\left< W_r \right> -A_{\tau}r^3 \right] = -R_c^* \,,
\end{equation}
A finite $R_c^*$ inhibit us from locating the peak of the RDF using (\ref{eqn_CR_diff_supp}) \`{a} la \cite{Lu10} i.e., without knowing $g(r)$, since $g(r)$ could no longer be factored out when $\frac{\partial g}{\partial r}=0$. However, we argue that (\ref{eqn_CR_diff_supp}) could still give a reasonably accurate account of the peak location. For the case of $St=0.05$, at $r =3d$ (the approximate peak location), we found the DNS data gives $-\tau_{\eta}B_{nl}\,r^4\frac{\partial g}{\partial r} \Big|_{\approx 0} + g(r)\left[ r^2\left< W_r \right> -A_{\tau}r^3 \right] \approx -1.05\!\times\!10^{-9} $ and $-R_c^* \approx -1.01\!\times\!10^{-9} $ %{\it (values should be revised with new gs(1)g0(1))} corroborating the preceding claim. 

\subsection{General Analytical Solution for the Differential Form of the Drift-Diffusion Equation.}

The general solution for the first-order non-homogenous ordinary differential equation (see, e.g., \cite{Arfken1999}), with $\left< w_r \right>_{\!*}$ given by the model in the main text, is:
\begin{equation} \label{gen_sol_supp}
	g(r) = \frac{1}{\beta(r)} \left[ \int \beta(r) q(r) dr + C \right] \,,
\end{equation} 
with $q(r) = R_c^*/(\tau_{\eta} B_{nl} r^4) $; $\beta(r) = \exp\left[ \int p(r)dr \right]$ and $p(r) = \left[ \, A_{\tau}r - \left< w_r \right>_{\!*} \, \right]/(\tau_{\eta} B_{nl} r^2) $. For the current model described in the main text, the integral in (\ref{gen_sol_supp}) could not be expressed in terms of simpler canonical functions. Hence, for specific applications, we currently anticipate  that some sort of power-law expansion or asymptotic reduction (if not numerical integration) would be needed to produce problem specific analytical approximations.

\subsection{Further Details on the Effects of Gravity.}

\begin{figure}
  \begin{center}
 	{\includegraphics[width=.45 \textwidth]{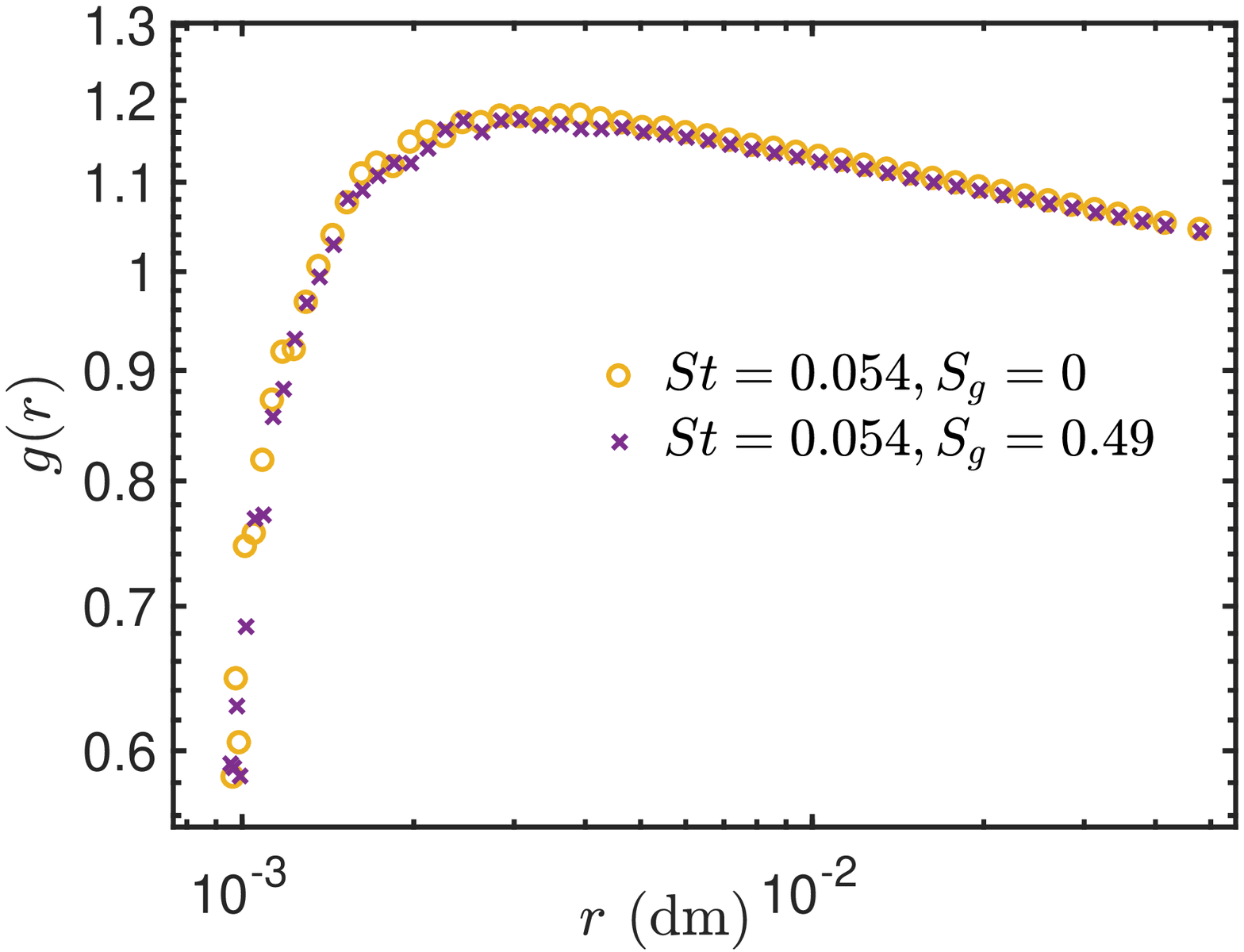}} \\ %\hspace{-20pt} 
     	%\vspace{-20.9pt}
     	{\includegraphics[width=.45 \textwidth]{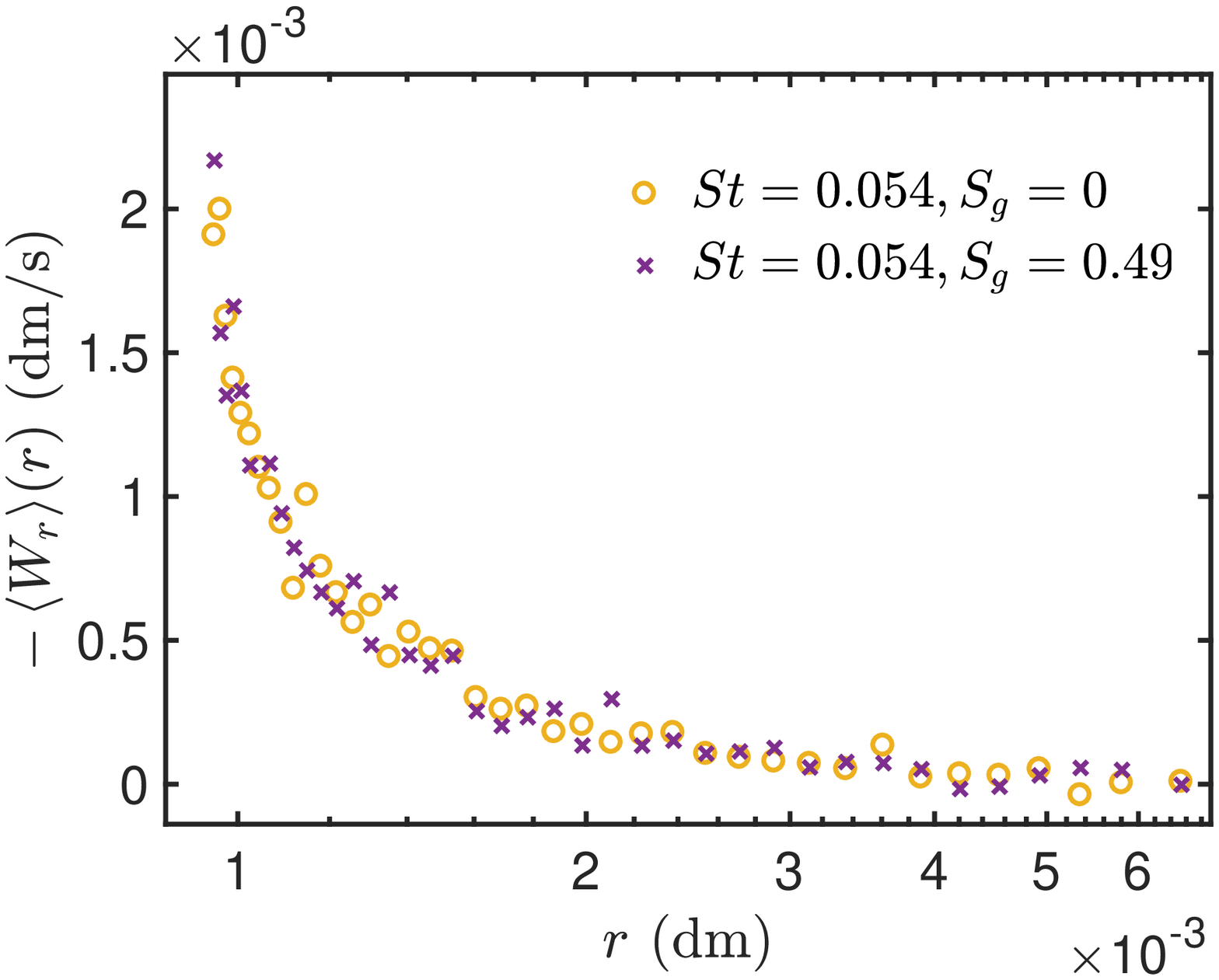}} %\hspace{-20pt}  \\ 
%     \begin{picture}(0,0)
%	\put(-80,80){\includegraphics[height=1.8cm]{Figures/RDF1.eps}}
%     \end{picture}
  \end{center}
\caption{\label{RDFnMRV_g_supp} {\bf Top)} RDFs of particles ($St=0.054$) subject to action of turbulence, collision-coagulation with and without gravity. Circles: $S_g=0$ (zero gravity); triangles: $S_g=0.49$ (nonzero gravity). No discernible effect of gravity. {\bf Bottom)} MRVs of the same cases. No discernible effect of gravity.}
\end{figure}

\ew{We repeat the DNS case of $St=0.054$ and $St=0.54$ with the particles subjected to gravity (body force), and compares results with the zero-gravity case. Fig.~\ref{RDFnMRV_g_supp} shows the results for case $St=0.054$. There is no discernible difference between the cases with and without gravity.}

\begin{figure}
  \begin{center}
 	{\includegraphics[width=.45 \textwidth]{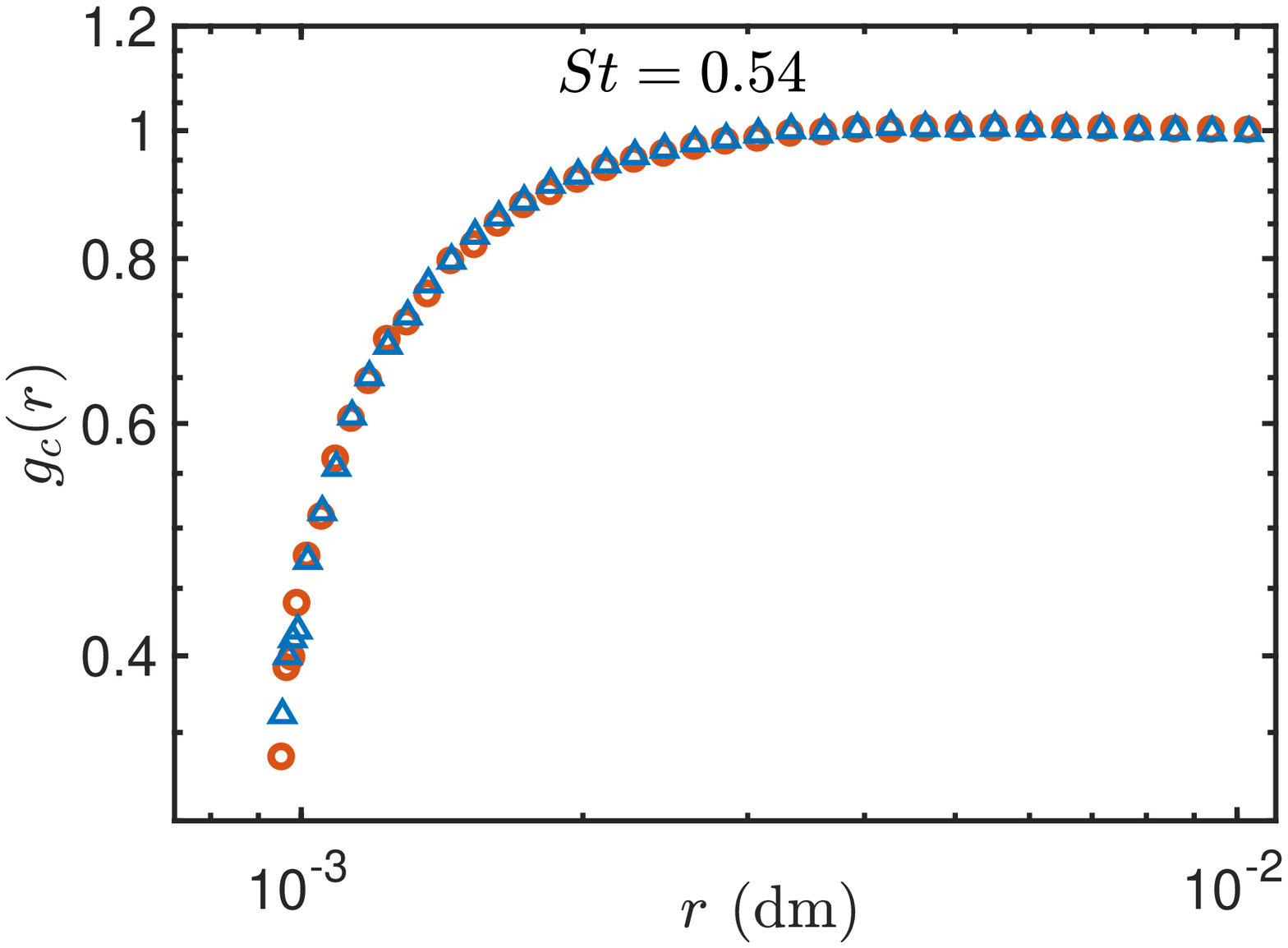}} \\ %\hspace{-20pt} 
     	%\vspace{-20.9pt}
%     \begin{picture}(0,0)
%	\put(-80,80){\includegraphics[height=1.8cm]{Figures/RDF1.eps}}
%     \end{picture}
  \end{center}
\caption{\label{RDF_comp_g_supp} Compensated RDFs of particles ($St=0.54$) subject to action of turbulence, collision-coagulation with and without gravity. Circles: $S_g=0$ (zero gravity); triangles: $S_g=4.9$ (nonzero gravity). Interpretation in the text.}
\end{figure}

\ew{For case $St=0.54$, the main RDF and MRV results is shown the main text. Here we show only the compensated-RDFs ( $g_c(r)$ ), where each $g_c(r)$ is calculated via $g(r)$ divide by a power law ($c_0 r^{-c_1}$) that resulted from curve-fitting to the original $g(r)$ in the range $0.6\eta \leq r \leq 3\eta$. Fig.~\ref{RDF_comp_g_supp} compares $g_c(r)$ for cases with and without gravity. The fact that there is no discernible difference implies that the \ews{uncompensated} $g(r)$ could be model as $g_c \times g_{\vec{g}}$ where $g_c$ is function that depends only on the particle collision process while $g_{\vec{g}}$ depends on other factors, e.g., gravity and is independent of particle collision. }

\end{document}